\documentclass[preprint2]{aastex}

\usepackage{graphicx}

\usepackage[authoryear]{}
\usepackage{longtable}

\newcommand\msun{M$_{\odot}$~}

\begin{document}

\title{The Occurrence of Non-Pulsating Stars in the $\gamma$ Dor and $\delta$ Sct Pulsation Instability Regions: \\Results from {\it Kepler} Quarter 14-17 Data\footnote{LaTeX typeset version of  \citet{guzik14} paper published in Astronomical Review, Vol. 10, No. 1, January 2015.}}

\author{J.A. Guzik,\altaffilmark{a} P.A. Bradley,\altaffilmark{b} J. Jackiewicz,\altaffilmark{c} J. Molenda-Zakowicz,\altaffilmark{c,d} K. Uytterhoeven\altaffilmark{e,f} and K. Kinemuchi\altaffilmark{g}}

\altaffiltext{a}{X-Theoretical Design Division, Los Alamos National Laboratory, XTD-NTA, MS T-086, Los Alamos, NM  87545 USA  e-mail: joy@lanl.gov}
\altaffiltext{b}{X-Computational Physics Division, Los Alamos National Laboratory, XCP-6, MS F-699, Los Alamos, NM  87545 USA}
\altaffiltext{c}{Department of Astronomy, New Mexico State University, Las Cruces, NM  88003  USA}
\altaffiltext{d}{Instytut Astronomiczny, Uniwersytet Wroclawski, Wroclaw, Poland}
\altaffiltext{e}{Instituto de Astrof{\`i}sica de Canarias (IAC), E-38200 La Laguna, Tenerife, Spain}
\altaffiltext{f}{Universidad de La Laguna, Dept. Astrof{\`i}sica, E-38206 La Laguna, Tenerife, Spain}
\altaffiltext{g}{Apache Point Observatory, Sunspot, NM 88349  USA}

\begin{abstract}
The high precision long time-series photometry of the NASA {\it Kepler} spacecraft provides an excellent means to discover and characterize variability in main-sequence stars, and to make progress in interpreting the pulsations to derive stellar interior structure and test stellar models.  For stars of spectral types A-F, the {\it Kepler} data revealed a number of surprises, such as more hybrid pulsating $\delta$ Sct and $\gamma$ Dor pulsators than expected, pulsators lying outside of the instability regions predicted by theory, and stars that were expected to pulsate, but showed no variability.

In our 2013 {\it Astronomical Review} article, we discussed the statistics of variability for 633 faint ({\it Kepler} magnitude 14-16) spectral type A-F stars observed by {\it Kepler} during Quarters 6-13 (June 2010-June 2012). We found six stars that showed no variability with amplitude 20 ppm or greater in the range 0.2 to 24.4 cycles/day, but whose positions in the log g-T$_{\rm eff}$ diagram place them in the $\delta$ Sct or $\gamma$ Dor pulsation instability regions established from pre-{\it Kepler} ground-based observations.  Here we present results for an additional 2137 stars observed during Quarters 14-17 (June 2012-May 2013). This sample is not unbiased, as we limited our target list to stars showing variability in Quarter 0 full-frame images to enhance our variable star discovery rate. We find that 990 stars, or 46\%, show no frequencies in the Fourier transform of their light curves down to the 20-ppm level, a smaller percentage than the 60\% of our Q6-13 sample.  We find 34 additional stars that lie within the ground-based $\gamma$ Dor/$\delta$ Sct pulsation instability regions; their lack of pulsations requires explanation.

In the analysis for our first paper, we included a +229 K offset to the {\it Kepler} Input Catalog T$_{\rm eff}$ to take into account an average systematic difference between the KIC values and the T$_{\rm eff}$ derived from SDSS color photometry for main-sequence F stars \citep{pinsonneault12, pinsonneault13}. We compare the KIC T$_{\rm eff}$ value and the T$_{\rm eff}$ derived from spectroscopy taken by the LAMOST instrument \citep{zakowicz13, zakowicz14} for 54 stars common to both samples.  We find no trend to support applying this offset; the trend instead shows that a small average temperature {\it decrease} relative to the KIC T$_{\rm eff}$ may be more appropriate for the stars in our spectral-type range. If the 229 K offset is omitted, only 17 of our 34 `constant' stars fall within the pulsation instability regions.  The comparisons with LAMOST-derived log g also show that the KIC log g may be too large for stars with KIC log g values $>$ 4.2.
For the two `constant' stars in the instability region that were also observed by LAMOST, the LAMOST T$_{\rm eff}$ values are cooler than the KIC T$_{\rm eff}$ by several hundred K, and would move these stars out of the instability regions.  It is possible that a more accurate determination of their T$_{\rm eff}$ and log g would move some of the other `constant' stars outside of the instability regions.  However, if average (random) errors in T$_{\rm eff}$ ($\pm$ 290 K) and an offset in log g from the KIC values are taken into account, 15 to 52 stars still persist within the instability regions.  Explanations for these `constant' stars, both theoretical and observational, remain to be investigated.

\end{abstract}


\section{Introduction}
\label{intro}

The NASA {\it Kepler} spacecraft was launched March 7, 2009 with the mission to detect transits of Earth-sized planets around sun-like stars using high precision CCD photometry \citep{borucki10}.  As a secondary mission, {\it Kepler} photometry has also been used to survey many thousands of stars for variability and stellar activity \citep{gilliland10}.
Through the NASA {\it Kepler} Guest Observer program Cycles 1-4, we requested and received observations of 2770 stars of spectral types A-F.  Our goals are to search for pulsating $\gamma$ Dor and $\delta$ Sct variable star candidates, identify `hybrid' stars pulsating in both types that are especially useful for asteroseismology and testing stellar pulsation theory, characterize the frequency content, and look for amplitude variability.   In the course of these observations, we discovered many eclipsing binary stars, stars that are probably not pulsating but show low-frequency photometric variations due to stellar activity or rotating star spots, and many stars that show no variability at the frequencies expected for $\gamma$ Dor or $\delta$ Sct stars.  An initial identification and categorization of the pulsating, binary, and spotted star candidates is given by \citet{bradley14, bradley15}.   A search for pulsations in one or more components of the eclipsing binaries in this sample has been started by \citet{gaulme14}. 

Paper 1 \citep{guzik13} discussed the number of `constant' stars, defined as having no frequencies with amplitude above 20 ppm in the frequency range of 0.2-24.4 cycles/day in long-cadence data on an unbiased sample of 633 stars observed during Quarters 6-13.  We found that about 60\% of the stars were `constant' by this definition.  We also identified six possible `constant' stars that lie within the boundaries of the $\delta$ Sct or $\gamma$ Dor pulsation instability regions as established by pre-{\it Kepler} ground-based observations.  This paper considers the `constant' stars in the larger not-unbiased sample of 2137 stars observed in long cadence during Q14-17, until the {\it Kepler} observations in the Cygnus-Lyra region ended after the failure of the spacecraft's second reaction wheel.

\section{Description of $\gamma$ Dor and $\delta$ Sct pulsators}

The $\gamma$ Doradus stars are late-A to F-type main sequence variables showing multiple periods of 0.3 to 3 days that have been identified as high-order, low-degree nonradial gravity (g) modes.  The intrinsic variability of the prototype $\gamma$ Doradus was discussed by \citet{balona94}, and the new class with 15 members was described by \citet{kaye99}.  Since then, over 60 members of the class have been identified \citep{henry07} using ground-based observations.  \citet{guzik00} proposed Ôconvective blockingÕ at the base of the envelope convective zones as the driving mechanism for $\gamma$ Dor g modes.  When the temperature at the convective envelope base is between 200,000 and 500,000 K, the convective timescale (mixing length/convective velocity) at this location is comparable to or longer than the g-mode pulsation period.  Since convection takes a portion of the pulsation cycle to turn on and transport the emergent luminosity, luminosity is periodically blocked, resulting in pulsation driving. This mechanism predicts unstable g modes in exactly the observed period range, and has become the accepted mechanism for $\gamma$ Dor pulsation driving \citep[see also][]{dupret04,dupret05,grigahcene06}.

The $\delta$ Scuti stars, on the other hand, pulsate in low-order radial and nonradial pressure (p) modes, although some of their modes have mixed p- and g-mode character; about 630 were catalogued as of 2000 \citep{rodriguez00}.  Their instability strip lies on the main sequence between early A and early F spectral types, and their periods range from just under 30 minutes to about 0.3 days.  Their pulsations are driven by the second ionization of helium that locally increases the opacity and regulates radiation flow in the envelope layers at $\sim$50,000 K \citep[the `$\kappa$ effect'; see, e.g.,][]{chevalier71}.

The $\gamma$ Dor and $\delta$ Sct stars are interesting asteroseismically because they form a bridge between the lower-mass solar-like stars with large convective envelopes and slow rotation, and the more massive, rapidly rotating stars with convective cores. The $\delta$ Sct and $\gamma$ Dor stars have both convective cores and convective envelopes, a variety of rotation rates, and live long enough for element settling, diffusion, and radiative levitation to alter their surface abundances and cause chemical peculiarities.  The envelope convection zones of the hotter and more massive $\delta$ Sct stars are confined to smaller regions (around 10,000-50,000 K) where hydrogen and helium are ionizing, and where convection transports a small fraction of the luminosity.  The envelope convection zones become deeper and carry a larger fraction of the star's luminosity with increasing stellar age and/or decreasing stellar mass \citep[see, e.g.,][and references therein]{turcotte98}.  Both $\gamma$ Dor and $\delta$ Sct stars pulsate in multiple modes that probe both the core and envelope.  The pulsations of these stars can be used to diagnose these phenomena and test stellar models.

Hybrid $\gamma$ Dor/$\delta$ Sct stars are among the most interesting targets for asteroseismology because the two types of modes (pressure and gravity) probe different regions of the star and are sensitive to the details of the two different driving mechanisms.  Because these driving mechanisms are somewhat mutually exclusive, hybrid stars exhibiting both types of pulsations are expected theoretically to exist only in a small overlapping region of temperature-luminosity space in the Hertzsprung-Russell (H-R) diagram \citep{dupret04, dupret05}.

Before the advent of the {\it Kepler} and CoRoT space missions, only four hybrid $\gamma$ Dor/$\delta$ Sct pulsators had been discovered. However, the first published analysis by the {\it Kepler} Asteroseismic Science Consortium (KASC) of 234 targets showing pulsations of either type revealed hybrid behavior in essentially all of them \citep{grigahcene10}.  In a study of 750 KASC A-F stars observed for four quarters \citep{uytterhoeven11}, 475 stars showed either $\delta$ Sct or $\gamma$ Dor variability, and 36\% of these were hybrids.  The {\it Kepler} hybrids were not confined to a small overlapping region of the two instability types in the temperature-luminosity space of the H-R diagram as predicted by theory.  Instead, they are found in the T$_{\rm eff}$-log g diagram throughout both the predicted $\gamma$ Dor and $\delta$ Sct instability regions, and even at cooler and hotter temperatures outside these regions.  Despite extensive study of this large sample and of the public data \citep{balonadziembowski11}, no obvious frequency or amplitude correlations with stellar properties have emerged, and there seems to be no clear observed separation of $\gamma$ Dor and $\delta$ Sct pulsators in the H-R diagram.  The known driving mechanisms cannot explain the pulsation behavior.

The existence and properties of these {\it Kepler} variable star candidates raise a number of questions:  Why are hybrids much more common than predicted by theory? Are additional pulsation driving mechanisms at work? Why are there apparently `constant' stars that lie within the instability regions but show no pulsation frequencies in the $\gamma$ Dor or $\delta$ Sct frequency range?

\section{Target selection and data processing}

We examined the {\it Kepler} data on the 2137 stars observed for {\it Kepler} Guest Observer Cycle 4 during Quarters 14 through 17 spanning the period June 28, 2012 to May 8,  2013.  These stars are relatively faint, with {\it Kepler} magnitudes 14-16.   We requested only long-cadence data, with integration time per data point of 29.4 minutes \citep{jenkins10}.  The Fourier transform of this data will be able to detect frequencies less than the Nyquist frequency, or half the sampling frequency, which is 24.4 cycles/day.  However, because $\gamma$ Dor stars have frequencies of about 1 cycle/day, and most $\delta$ Sct stars have frequencies of 10-20 cycles/day, the long-cadence data is adequate to identify $\gamma$ Dor and most $\delta$ Sct candidates.  Also, as discussed below, super-Nyquist frequencies will appear mirrored at sub-Nyquist frequencies \citep{murphy13} and therefore will be detected in the amplitude spectrum, if they exist.

We chose these stars by using the {\it Kepler} Guest Observer target selection tool to search the {\it Kepler} Input Catalog \citep[KIC,][]{latham05}.  For Q6-13 discussed in Paper 1, we restricted our sample to stars in or near the $\gamma$ Dor/$\delta$ Sct instability regions, with effective temperature 6200 $<$ T$_{\rm eff}$ $<$ 8300 K; surface gravity 3.6 $<$ log g $<$ 4.7, contamination or blend from background stars $<$ 0.01, and {\it Kepler} Flag 0, meaning that these targets had not yet been observed by {\it Kepler} .  Most of the brighter stars had already been observed by either the {\it Kepler} Science Team, or KASC, and these are discussed by \citet{uytterhoeven11}. For Quarter 14-17 targets observed for our Cycle 4 Guest Observer proposal reported here, the selection was not unbiased, but instead was cross-correlated with a list of potentially variable stars identified by comparing full-frame images taken during the {\it Kepler} spacecraft commissioning period \citep{kinemuchi11}.   In our target search, we shifted our criteria to lower temperatures (5900-8000 K), as we were finding many stars with $\gamma$ Dor and $\delta$ Sct pulsations at lower T$_{\rm eff}$.  We also relaxed the contamination factor criteria to $<$ 0.05.  This search yielded a sample of over 6000 stars that we cross-correlated against the $\sim$300,000 stars in the {\it Kepler} field found to be variable by \citet{kinemuchi11} by comparing the Q0 full-frame images.  The cross-correlation returned 2174 stars, of which we received data for 2137 stars

After downloading the data files from the MAST (Mikulski Archive for Space Telescopes, http://archive.stsci.edu/index.html), we processed them using Matlab scripts developed by J. Jackiewicz.  We combined data for all available quarters for each star.  We chose to use the data corrected by the {\it Kepler} pipeline; we have also tried using the raw light curves, and found essentially no difference in the frequencies found by the Fourier transform.  

The Matlab scripts remove outlier points and interpolate the light curves to an equidistant time grid.  The light curves are then converted from {\it Kepler} flux (FK) to parts per million (ppm) using the formula f(t) = 10$^6$(F$_K$/y - 1), where y is either the mean value of the entire light curve, or a low-order polynomial fit to the light curve, depending on artifacts present in the data \citep[see also][]{mcnamara12}.  After the processing and Fourier transform of the data, the Matlab scripts generate a plot for each star, such as the examples shown in Figs. \ref{CSFig2a} through \ref{CSFig2f}.  In these plots, the bottom panel is the light curve in ppm vs. time; the top panel shows the amplitude in ppm vs. frequency in cycles/day determined by the Fourier transform, and the middle two panels are enlargements of the frequency ranges 0-5 cycles/day and 5-24.4 cycles/day.  The header on each panel shows the KIC number, the Quarters of observation, the {\it Kepler} magnitude, and log surface gravity (log g, rounded to two digits).  The radius in the header is also from the KIC but is not accurate, as it is derived by assuming a stellar mass of 1 \msun and using the log surface gravity (log g) of the KIC.

\section{Identification and distribution of `constant' stars in the H-R diagram}

As in Paper 1, we examined individually the Matlab-generated plots for our sample of 2137 stars.  We flagged as `constant' the stars with no power in the spectrum from 0.2 to 24.4 c/d above 20 ppm.  

There is some judgment involved in deciding whether a given star meets these criteria.  For some stars, the noise level over a good portion of this frequency range is about 20 ppm.  For many stars, stellar activity may cause photometric variations and power in the spectrum at slightly above the 20 ppm level at 0.2 to 0.5 c/d frequencies.  We eliminated from consideration stars that appear to be eclipsing binaries that may have stellar components that individually turn out to be either `constant' or variable once the binary signal is removed.  We have begun analysis of the binaries \citep{gaulme14}.

\begin{figure*}%tb
\center
\includegraphics[width=1.5\columnwidth]{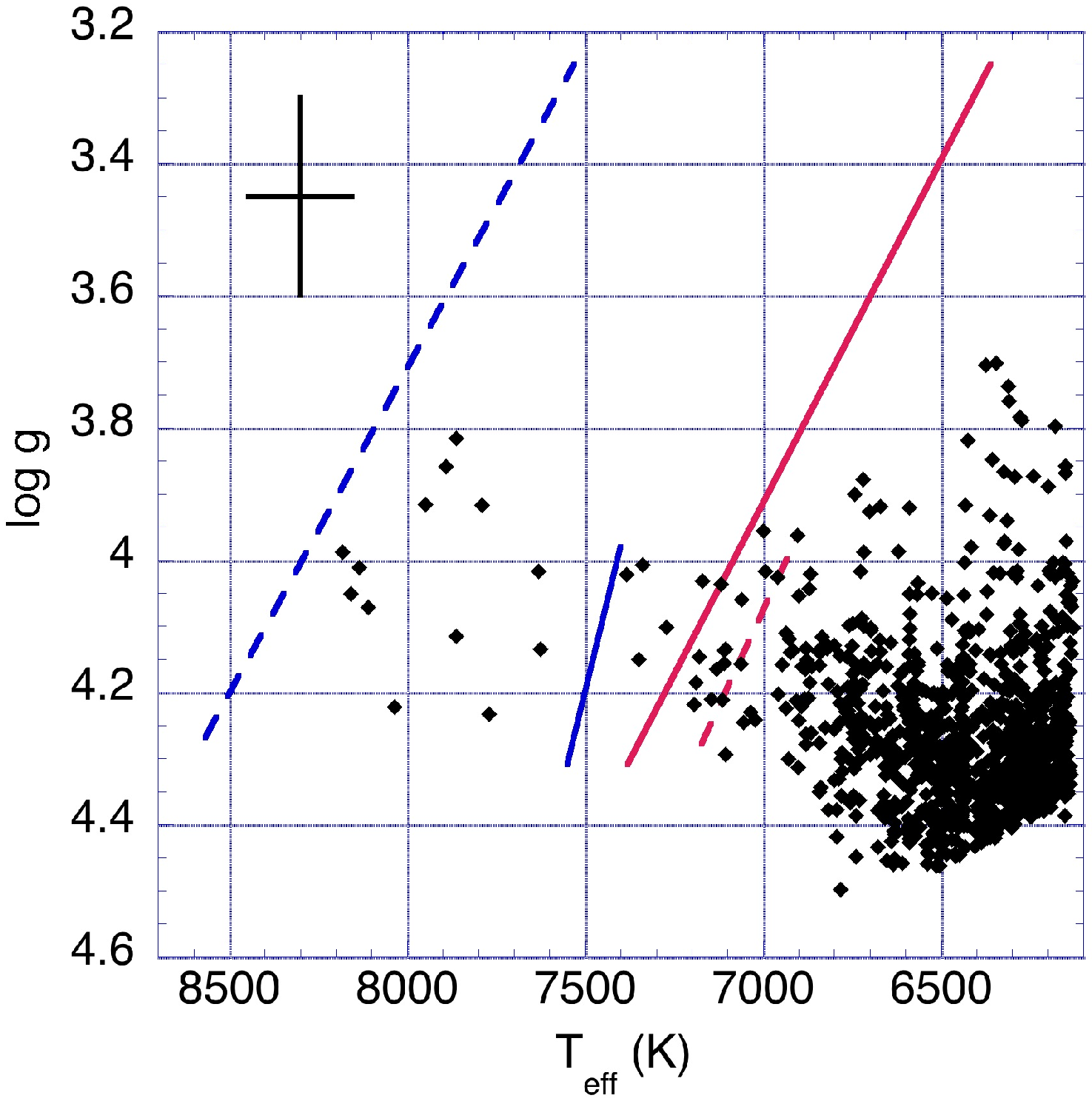}
\caption{Location of sample stars that are `constant' (see definition in text) in the log surface gravity  - T$_{\rm eff}$ diagram, along with $\delta$ Sct (dashed lines) and $\gamma$ Dor (solid lines) instability strip boundaries established from pre-{\it Kepler} ground-based observations \citep{rodriguezbreger01, handlershobbrook02}.  The T$_{\rm eff}$ of the sample stars has been shifted by +229 K to account for the systematic offset between T$_{\rm eff}$ of the {\it Kepler} Input Catalog and SDSS photometry for this temperature range as determined by \citet{pinsonneault12}.  The black cross shows an error bar on log g (0.3 dex) and T$_{\rm eff}$ (290 K) established by comparisons of KIC values and values derived from ground-based spectroscopy for brighter {\it Kepler} targets \citep{uytterhoeven11}.}
\label{CSFig1a}
\end{figure*}

In Paper 1, by these criteria we found 359 `constant' stars out of the 633 stars in our sample, or roughly 60\%.   Here, with this preconditioned sample, we found 990 out of 2137 stars, or 46\% that are `constant'.  We plotted the position of the `constant' stars in a log surface gravity vs. T$_{\rm eff}$ diagram (\ref{CSFig1a}), along with the $\delta$ Sct (dashed lines) and $\gamma$ Dor (solid lines) instability strip boundaries established from pre-{\it Kepler} ground-based observations \citep{rodriguezbreger01, handlershobbrook02}.We show with a black cross the error bar on log g (0.3 dex) and T$_{\rm eff}$ (290 K) derived by comparing KIC values of brighter {\it Kepler} targets with those derived from ground-based spectroscopy, as discussed by \citet{uytterhoeven11}.

As in Paper 1, we applied a 229 K increase to the effective temperature given by the KIC. This shift is used to roughly account for the systematic offset between temperatures given in the KIC and Sloan Digital Sky Survey photometry for the temperature range of our stellar sample as determined by \citet{pinsonneault12, pinsonneault13}.  To determine this shift, we averaged the five temperature offsets given in table 8 of \citet{pinsonneault12} for effective temperatures 6200-6600 K.  The 1-sigma error in these offsets is about 35 K.  

\begin{figure*}%tb
\center
\includegraphics[width=1.5\columnwidth]{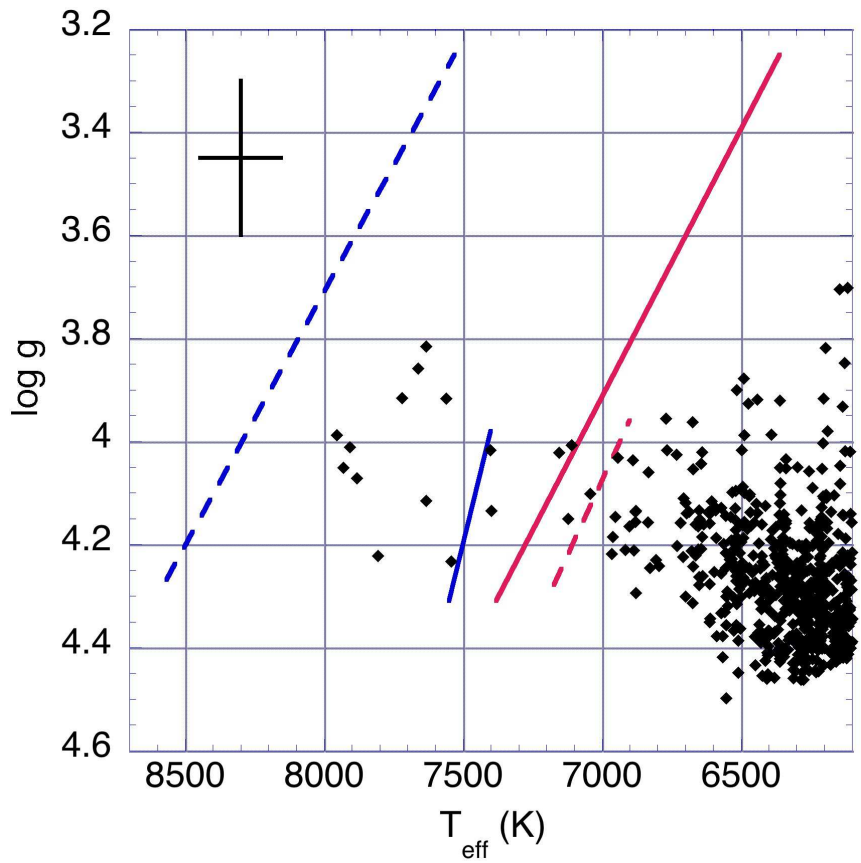}
\caption{Same as Fig. \ref{CSFig1a}, but without the +229 K offset applied to the KIC T$_{\rm eff}$ values.  Without the offset, only 17 stars lie within the pulsation instability regions.}
\label{CSFig1b}
\end{figure*}

Note that the instability strip boundaries in Figs. \ref{CSFig1a} and \ref{CSFig1b} were derived from ground-based pre-{\it Kepler} observations of known $\gamma$ Dor and $\delta$ Sct stars with relatively accurate determinations of T$_{\rm eff}$ and log g from multicolor photometry or spectroscopy.  The instability region boundaries derived from theoretical stellar evolution models depend on many factors, including helium and element abundance, abundance mixtures, and convection treatment.  The instability region boundaries predicted by theory were in relatively good agreement with the ground-based observations prior to the {\it Kepler} data; however, the distribution of the many new $\delta$ Sct and $\gamma$ Dor candidates studied by KASC indicate that the instability strip boundaries may be wider than predicted by theoretical models and defined by pre-{\it Kepler} ground-based observations \citep[see, e.g.,][]{uytterhoeven11}, or that additional pulsation instability mechanisms may need to be considered to explain the data.

Figure \ref{CSFig1a} shows that 34 of the 990 `constant' stars, or only 1.6\% of the entire sample, lie within the instability strip boundaries derived from the ground-based variable star observations.  The uncertainties on T$_{\rm eff}$ and log g for our sample may mean that some of the stars in our sample are not in the instability regions, but just as easily, some of the stars just outside the instability region may lie within the instability region.  Table \ref{table1} lists the 34 stars, along with their {\it Kepler} magnitude, KIC log g and KIC T$_{\rm eff}$ with and without the +229 K offset.  Figure \ref{CSFig1b} shows the same diagram as in Fig. \ref{CSFig1b}, except without the +229 K offset applied to the KIC T$_{\rm eff}$ values.  If the offset is not applied, only 17 `constant' stars remain in the instability regions. 

\begin{figure*}%tb
\center
\includegraphics[width=1.5\columnwidth]{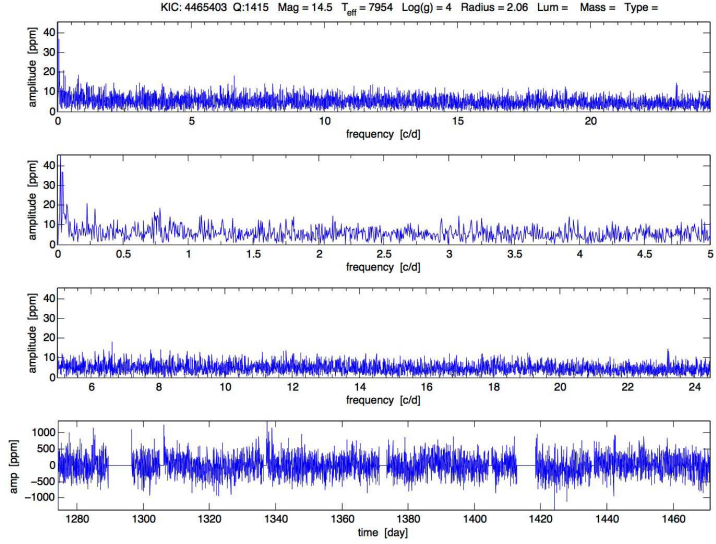}
\caption{Amplitude spectrum (top 3 panels) and light curve (bottom panel) for KIC 4465403, the hottest of the `constant' stars within the $\delta$ Sct instability strip boundary in Figs. \ref{CSFig1a} and \ref{CSFig1b}.  The noise level for this faint star is at about the 10 ppm level.}
\label{CSFig2a}
\end{figure*}

\begin{figure*}%tb
\center
\includegraphics[width=1.5\columnwidth]{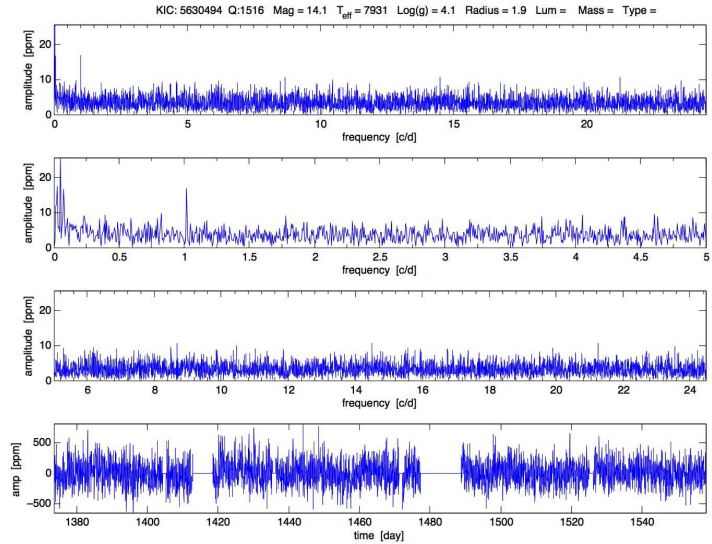}
\caption{Amplitude spectrum (top 3 panels) and light curve (bottom panel) for KIC 5630494, another hot `constant' star within the $\delta$ Sct instability region in Figs. \ref{CSFig1a} and \ref{CSFig1b}.  The noise level is about 7 ppm.  One probably non-significant frequency at about 1 c/d is evident with amplitude about 17 ppm.}
\label{CSFig2b}
\end{figure*}

\begin{figure*}%tb
\center
\includegraphics[width=1.5\columnwidth]{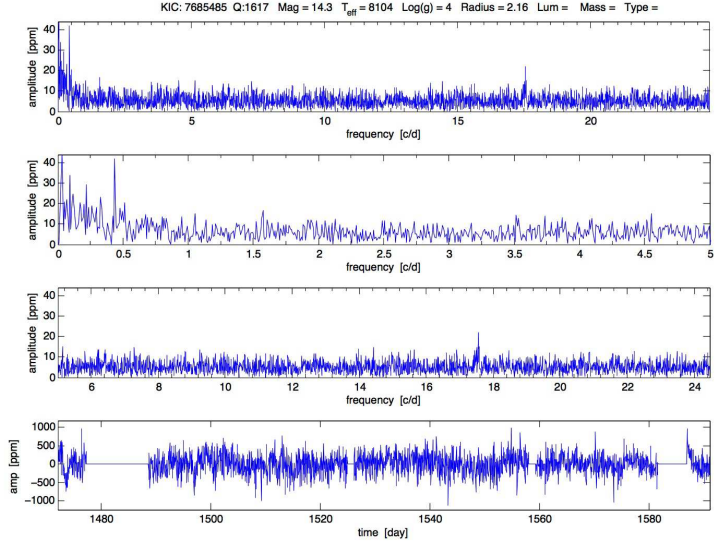}
\caption{Amplitude spectrum (top 3 panels) and light curve (bottom panel) for KIC 7685485, another hot `constant' star within the $\delta$ Sct instability region in Figs. \ref{CSFig1a} and \ref{CSFig1b}.  The noise level is about 10 ppm.  One frequency at about 0.4 c/d with amplitude of 40 ppm is evident, but a longer data set is needed to confirm.  The frequency at about 17.5 c/d is present for many of the stars in the Q16-Q17 data sets and is an artifact.}
\label{CSFig2c}
\end{figure*}

\begin{figure*}%tb
\center
\includegraphics[width=1.5\columnwidth]{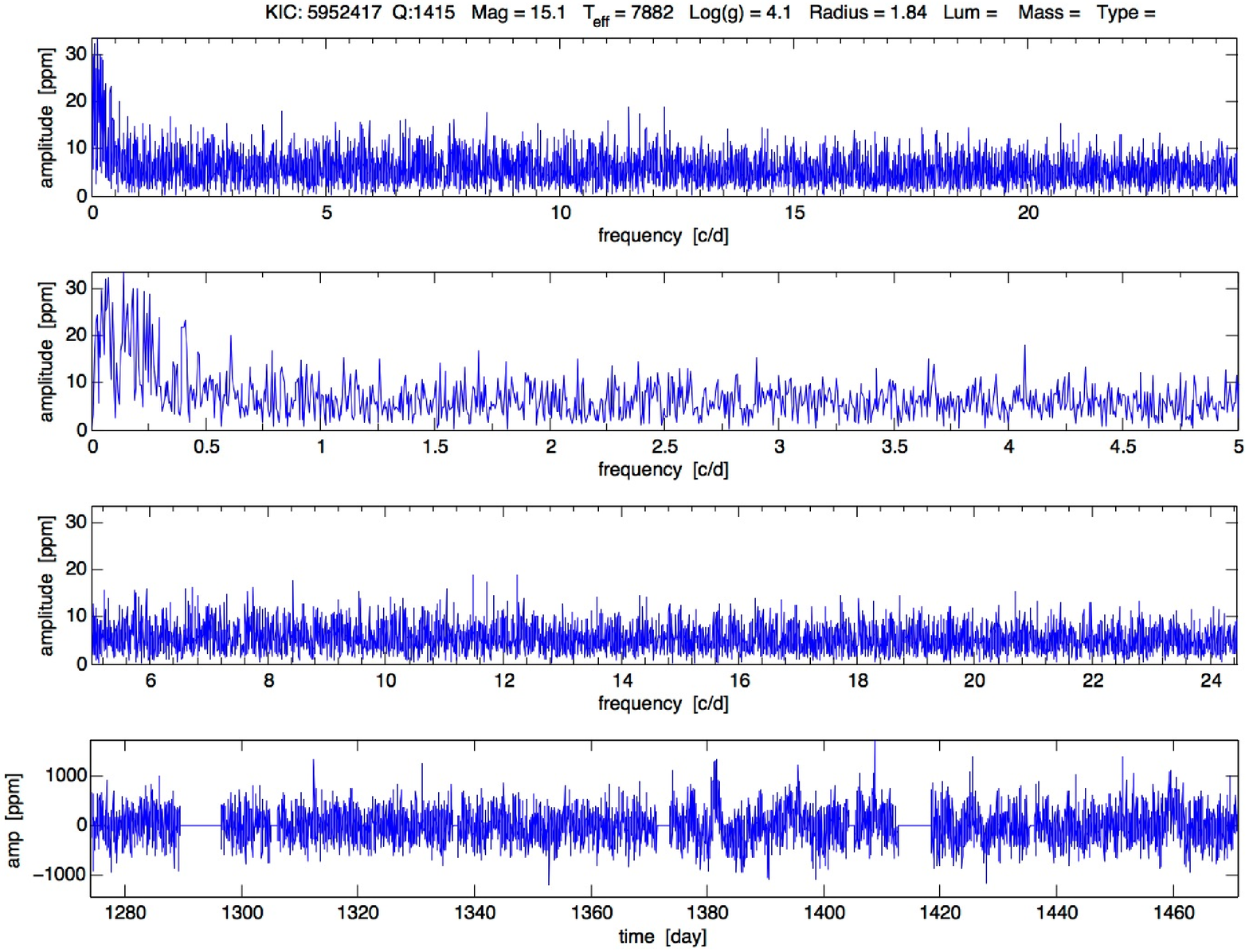}
\caption{Amplitude spectrum (top 3 panels) and light curve (bottom panel) for KIC 5952417, another hot `constant' star within the $\delta$ Sct instability region in Figs. \ref{CSFig1a} and \ref{CSFig1b}.  The noise level is about 10 ppm.}
\label{CSFig2d}
\end{figure*}

\begin{figure*}%tb
\center
\includegraphics[width=1.5\columnwidth]{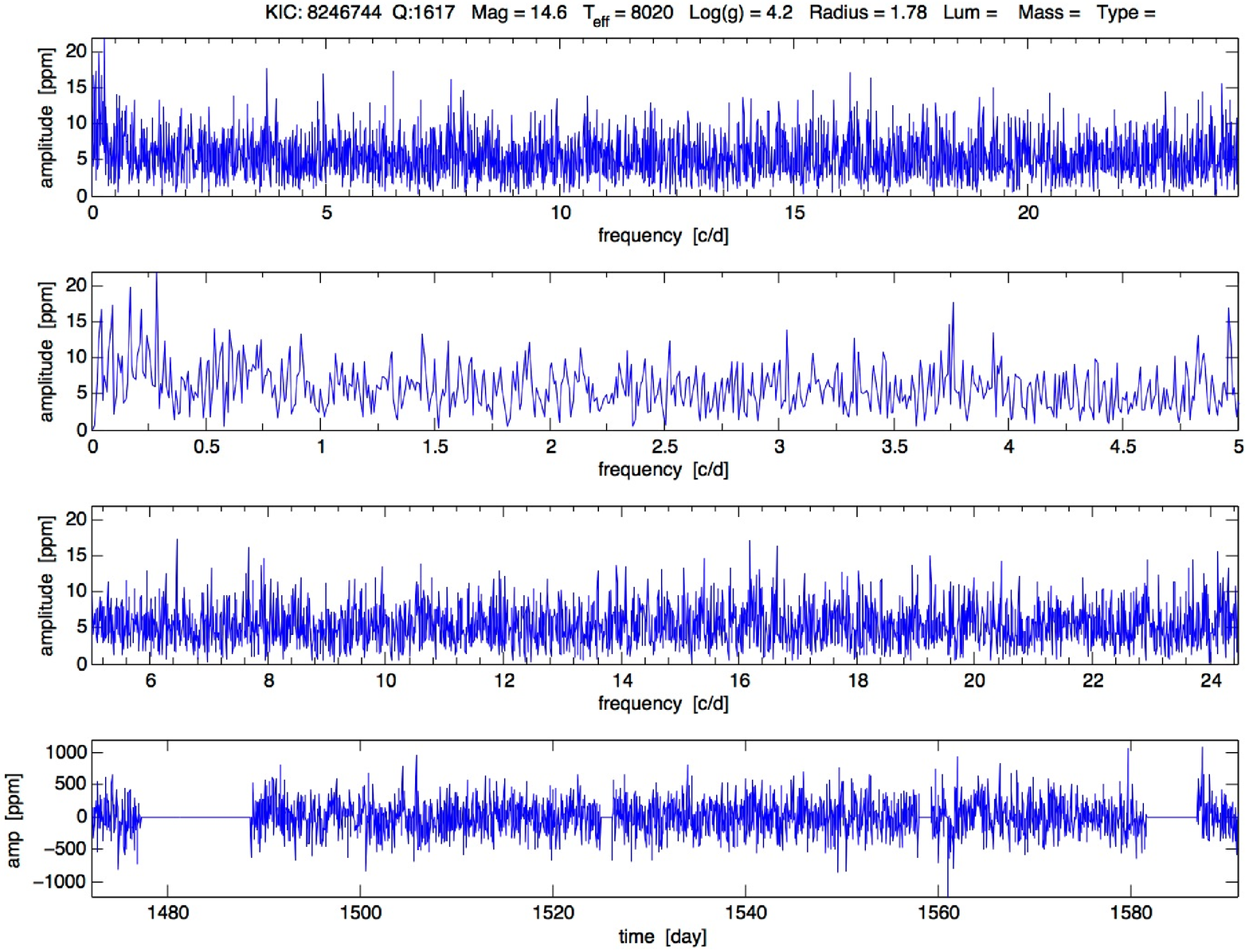}
\caption{Amplitude spectrum (top 3 panels) and light curve (bottom panel) for KIC 8246744, another hot `constant' star within the $\delta$ Sct instability region in Figs. \ref{CSFig1a} and \ref{CSFig1b}.  The noise level is about 10 ppm.}
\label{CSFig2e}
\end{figure*}

\begin{figure*}%tb
\center
\includegraphics[width=1.5\columnwidth]{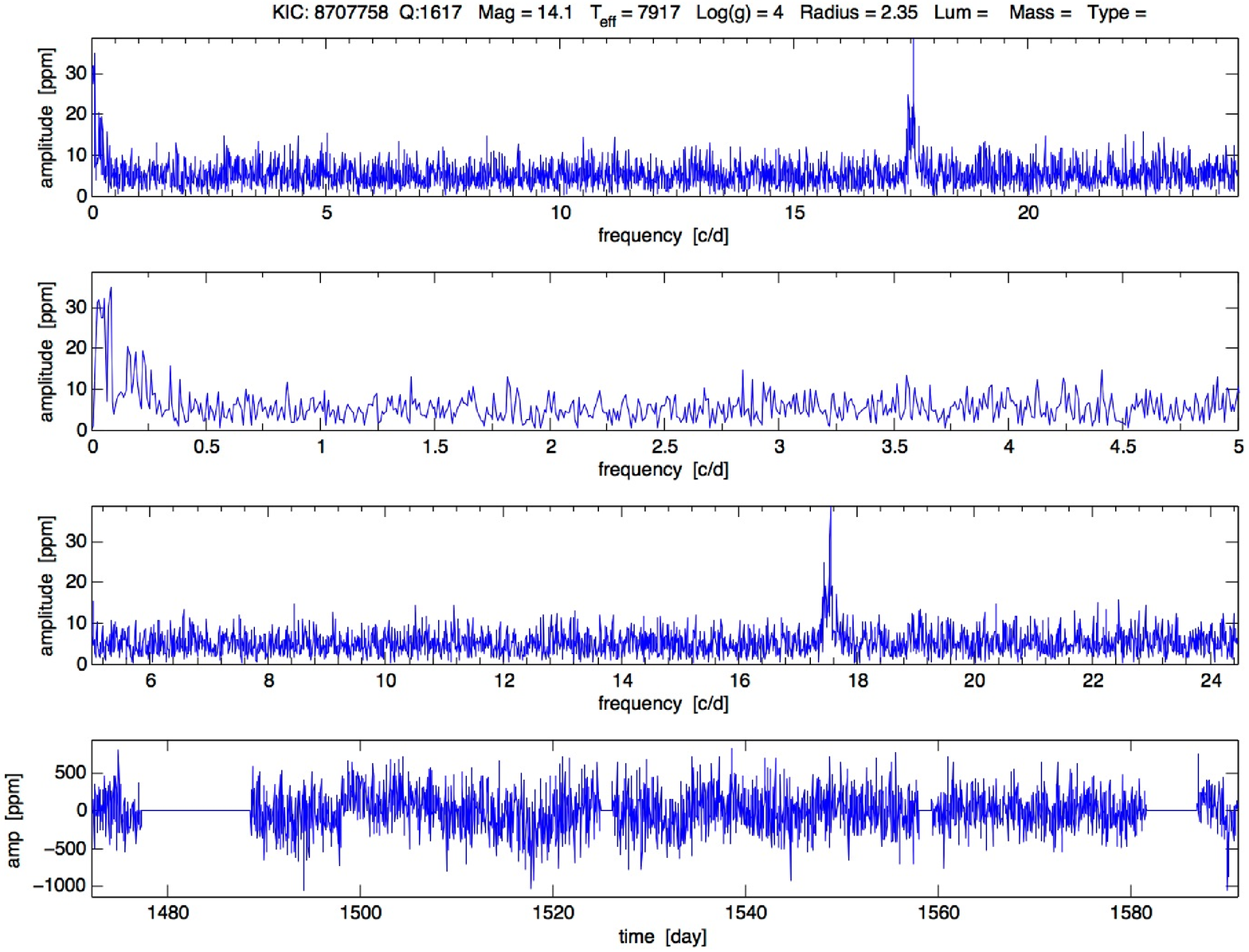}
\caption{Amplitude spectrum (top 3 panels) and light curve (bottom panel) for KIC 8707758, another hot `constant' star within the $\delta$ Sct instability region in Figs. \ref{CSFig1a} and \ref{CSFig1b}.  The noise level is about 10 ppm; the artifact frequency $\sim$17.5 c/d also evident in Fig. \ref{CSFig2b} is clearly seen.}
\label{CSFig2f}
\end{figure*}

We show the light curves and amplitude spectra from the Fourier transforms for six of the hottest of these stars Figs. \ref{CSFig2a} through \ref{CSFig2f} that will lie within the pulsation instability regions of Figs. \ref{CSFig1a} and \ref{CSFig1b} whether or not the +229 K T$_{\rm eff}$ offset is applied.  The figure captions include comments on the light curve and amplitude spectrum details.                               

\begin{deluxetable}{ccccc}
\tablecolumns{5}
\tablewidth{0pt}
% \begin{minipage}{150mm}
\tablecaption{KIC Number, effective temperature, log g and {\it Kepler} magnitude for 34 `constant' stars that lie within the pre-{\it Kepler} pulsation instability regions.}
\tablehead{\colhead{KIC Number}      &
    \colhead{KIC T$_{\rm eff}$ (K)}             &
    \colhead{KIC T$_{\rm eff}$ + 229 K}             &
    \colhead{KIC log g}                   &
    \colhead{K$_p$ mag}               }
\startdata
4465403  &  7954  &  8183  &  3.988  &  14.498\\
5630494  &  7931  &  8160  &  4.051  &  14.078\\
7685485  &  7907  &  8136  &  4.011  &  14.339\\
5952417  &  7882  &  8111  &  4.071  &  15.064\\
8246744  &  7807  &  8036  &  4.222  &  14.622\\
8707758  &  7720  &  7949  &  3.916  &  14.091\\
7620739  &  7663  &  7892  &  3.858  &  14.621\\
6139295  &  7634  &  7863  &  4.115  &  14.374\\
7467076  &  7634  &  7863  &  3.816  &  14.67\\
4661844  &  7562  &  7791  &  3.917  &  14.921\\
6279284  &  7543  &  7772  &  4.233  &  14.661\\
7595223  &  7402  &  7631  &  4.017  &  14.442\\
9590213  &  7398  &  7627  &  4.135  &  14.432\\
8580178  &  7155  &  7384  &  4.022  &  14.597\\
5096805  &  7122  &  7351  &  4.15  &  14.51\\
4171238  &  7110  &  7339  &  4.007  &  14.227\\
3729981  &  7043  &  7272  &  4.101  &  14.084\\
9016039  &  6966  &  7195  &  4.218  &  14.267\\
8355866  &  6961  &  7190  &  4.185  &  14.349\\
7751249  &  6952  &  7181  &  4.146  &  14.066\\
7668077  &  6943  &  7172  &  4.031  &  14.34\\
6522027  &  6916  &  7145  &  4.21  &  14.291\\
6696719  &  6902  &  7131  &  4.165  &  14.274\\
7517640  &  6890  &  7119  &  4.036  &  14.767\\
6530559  &  6886  &  7115  &  4.211  &  14.912\\
6124552  &  6880  &  7109  &  4.156  &  14.33\\
8292900  &  6880  &  7109  &  4.136  &  14.307\\
7364349  &  6878  &  7107  &  4.135  &  14.421\\
5950897  &  6835  &  7064  &  4.157  &  14.401\\
6035807  &  6833  &  7062  &  4.06  &  15.413\\
5089044  &  6771  &  7000  &  3.956  &  14.594\\
4375039  &  6766  &  6995  &  4.017  &  14.558\\
3444020  &  6732  &  6961  &  4.026  &  14.727\\
4271591  &  6675  &  6904  &  3.962  &  15.00\\
\enddata
\label{table1}
\end{deluxetable}  

\section{T$_{\rm eff}$ and log g determinations from LAMOST spectra}

Because our sample of {\it Kepler} stars is faint (K$_p$ mag 14-16), temperatures, metallicities, and surface gravities obtained spectroscopically were not available when the targets were selected.  The LAMOST (Large Sky Area Multi-Object Fiber Spectroscopic Telescope) project \citep{zakowicz13, zakowicz14} is in the process of obtaining and analyzing spectra of thousands of stars in the {\it Kepler} field-of-view.  As of this writing, 54 of our 2137 stars had been observed with LAMOST; a few had more than one spectra taken and reduced months apart.  Table \ref{table2} compares the T$_{\rm eff}$ and log g derived from the LAMOST spectra with the KIC T$_{\rm eff}$ and log g values for these 54 stars.  The uncertainties reflect the internal precision of the spectroscopic reduction method; the actual uncertainties may be higher.

\begin{figure*}%tb
\center
\includegraphics[width=1.5\columnwidth]{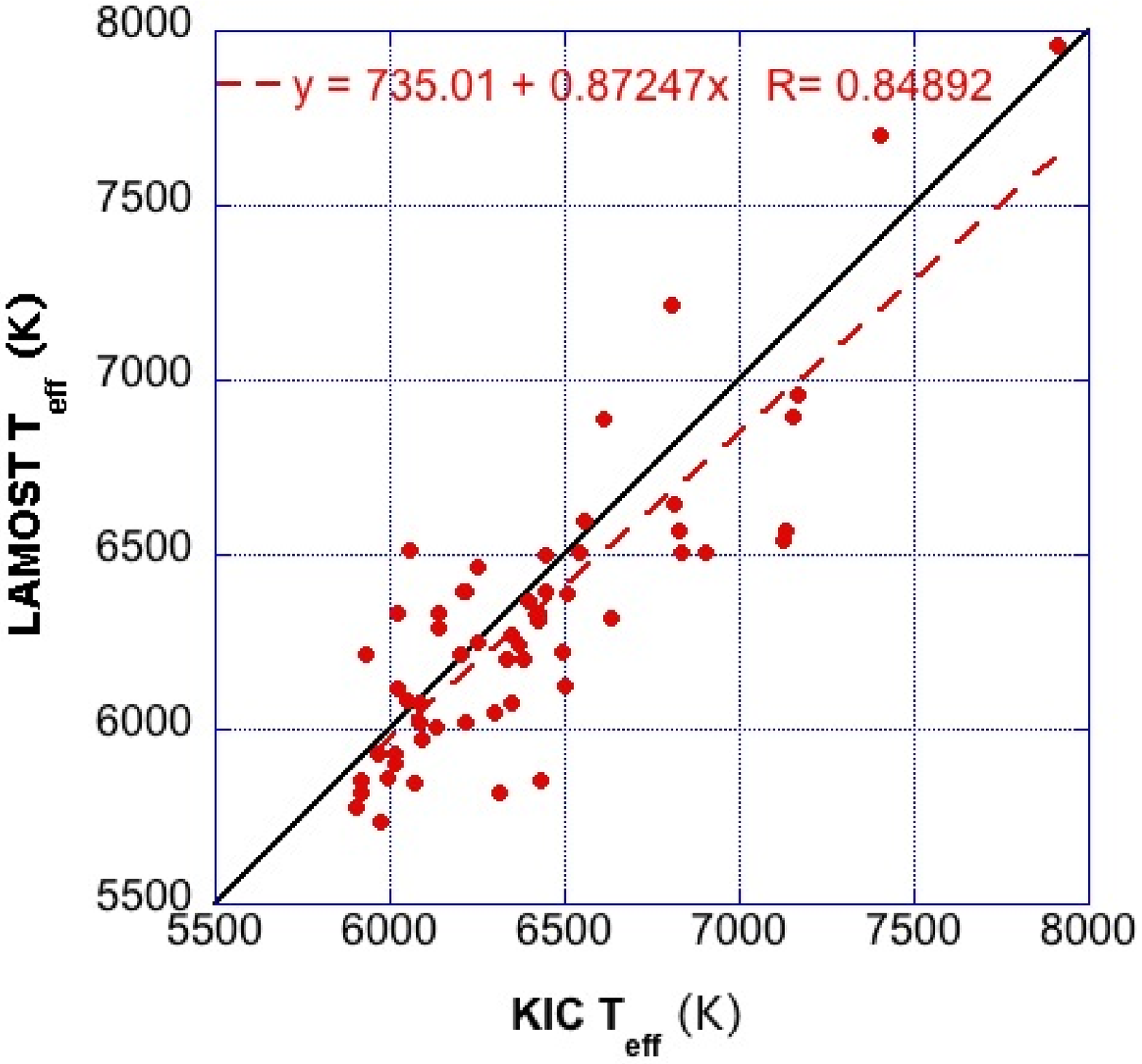}
\caption{LAMOST T$_{\rm eff}$ vs. KIC T$_{\rm eff}$ for 54 stars with T$_{\rm eff}$ derived from LAMOST spectra. Three stars have two different spectra and T$_{\rm eff}$ derivations.  The black line (slope = 1) shows the trend if the temperatures were in agreement.  The red line shows a least-squares fit to the data (slope = 0.87, correlation coefficient 0.85), indicating the spectroscopically derived T$_{\rm eff}$ are on average slightly lower than those of the KIC T$_{\rm eff}$, opposite in sign to the average corrections found by \citet{pinsonneault12} from Sloan photometry-derived T$_{\rm eff}$ for stars of these A-F spectral types.}
\label{CSFig3a}
\end{figure*}

\begin{figure*}%tb
\center
\includegraphics[width=1.5\columnwidth]{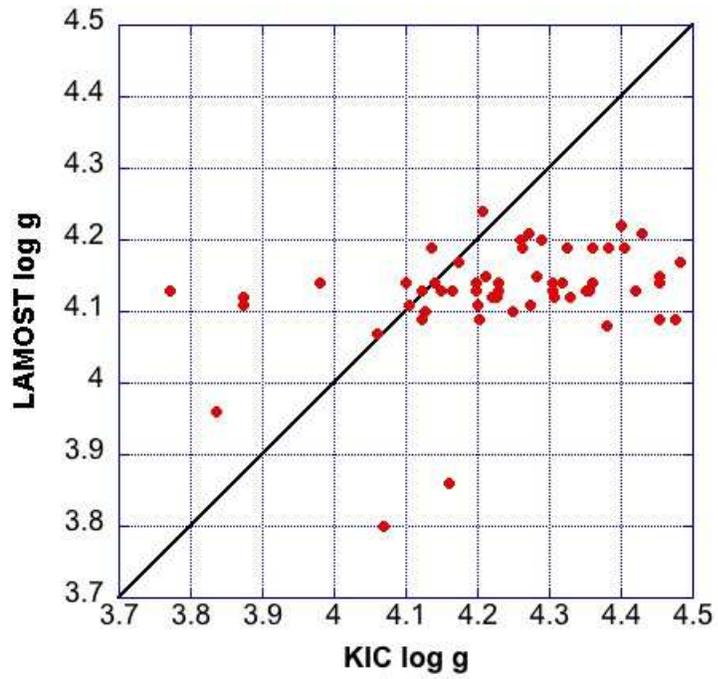}
\caption{LAMOST log g vs. KIC log g for 54 stars with T$_{\rm eff}$ derived from LAMOST spectra. Three stars have two different spectra and T$_{\rm eff}$ derivations.  The black line shows the trend if the log g values were in agreement.  The LAMOST log g values plateau at values of about 4.2 for stars of these spectral types, whereas the KIC values are spread to values as high as 4.5.}
\label{CSFig3b}
\end{figure*}

Figures \ref{CSFig3a} and \ref{CSFig3b} show the KIC T$_{\rm eff}$ and log g plotted against the values derived from LAMOST spectra.  The LAMOST T$_{\rm eff}$ values are sometimes higher, and sometimes lower than the KIC values, with the difference typically $\sim$200 K.   A least-squares fit shows that the LAMOST-derived T$_{\rm eff}$ values tend to be slightly lower on average than the KIC values, opposite to the average difference determined by the study of \citet{pinsonneault12} using SDSS color photometry.  The plot of LAMOST-derived log g vs. KIC log g shows that the log g reaches a plateau at about 4.2, whereas the KIC log g continues to increase to 4.5.  An average overestimation bias of about 0.23 dex in KIC log g was noted by \citet{verner11}.

\begin{figure*}%tb
\center
\includegraphics[width=1.5\columnwidth]{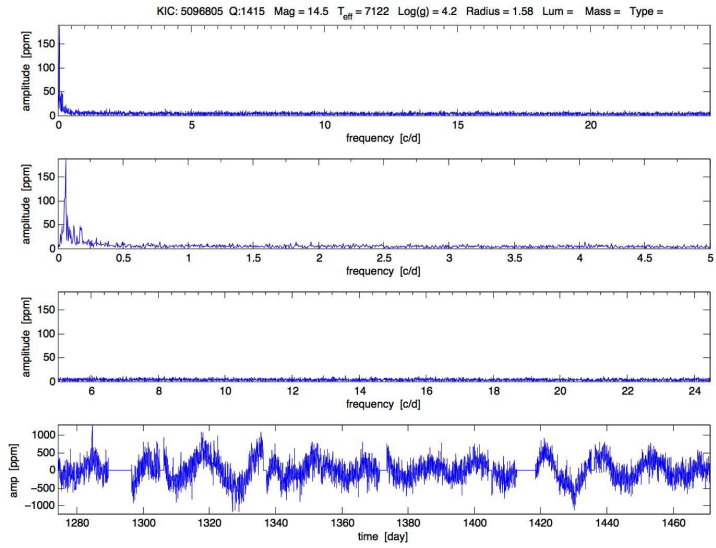}
\caption{KIC 5096805 light curve (bottom panel) and amplitude spectrum (top panels), categorized as a `constant' star by our amplitude spectrum criteria.  The light curve variations are indicative of starspots rotating in and out of visibility and changing in size.  The LAMOST-derived T$_{\rm eff}$ for this star is almost 600 K cooler than the KIC T$_{\rm eff}$ value of 7122 K, and, if adopted, would place it outside of the pulsation instability region (Fig. \ref{CSFig5b}).}
\label{CSFig4a}
\end{figure*}

\begin{figure*}%tb
\center
\includegraphics[width=1.5\columnwidth]{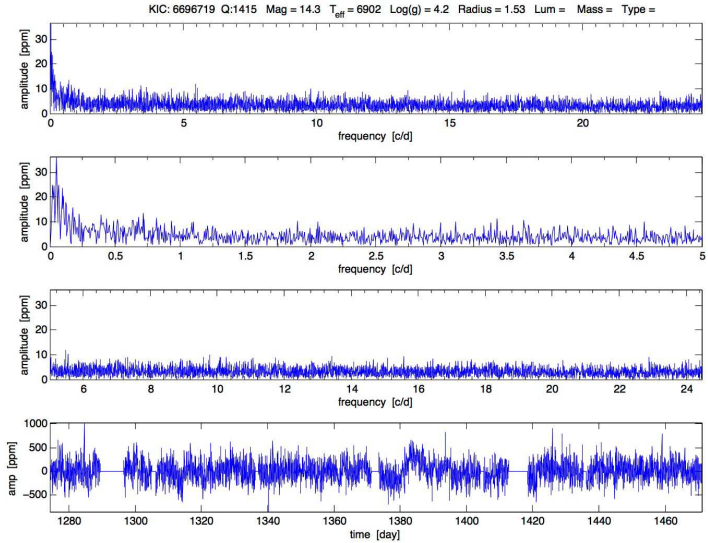}
\caption{KIC 6696719 light curve (bottom panel) and amplitude spectrum (top panels), categorized as a `constant' star. The LAMOST-derived T$_{\rm eff}$ for this star is almost 400 K cooler than the KIC T$_{\rm eff}$ value of 6902 K, and, if adopted, would place this star outside of the pulsation instability region (Fig. \ref{CSFig5b}).}
\label{CSFig4b}
\end{figure*}

\begin{figure*}%tb
\center
\includegraphics[width=1.5\columnwidth]{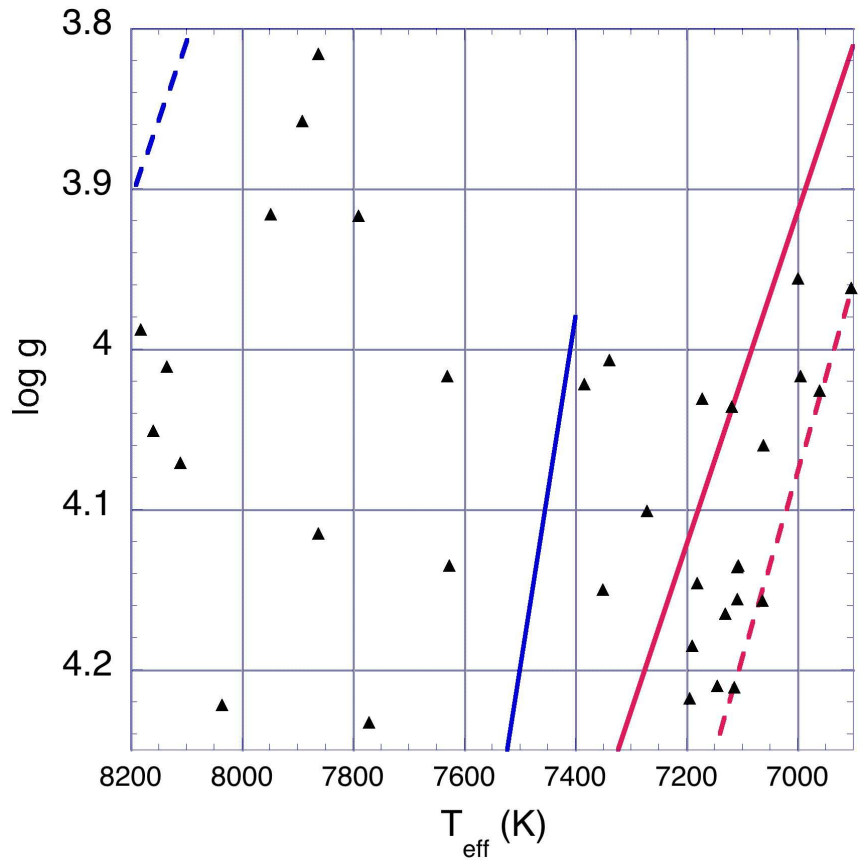}
\caption{Enlargement of pulsation instability region of Fig. \ref{CSFig1a} including the +229 K offset from KIC T$_{\rm eff}$ values.  The 34 `constant' stars of Table \ref{table1} are shown here.  The symbols for two stars with nearly the same T$_{\rm eff}$ and log g (KIC 7364349 and KIC 8292900) overlap.}
\label{CSFig5a}
\end{figure*}

\begin{figure*}%tb
\center
\includegraphics[width=1.5\columnwidth]{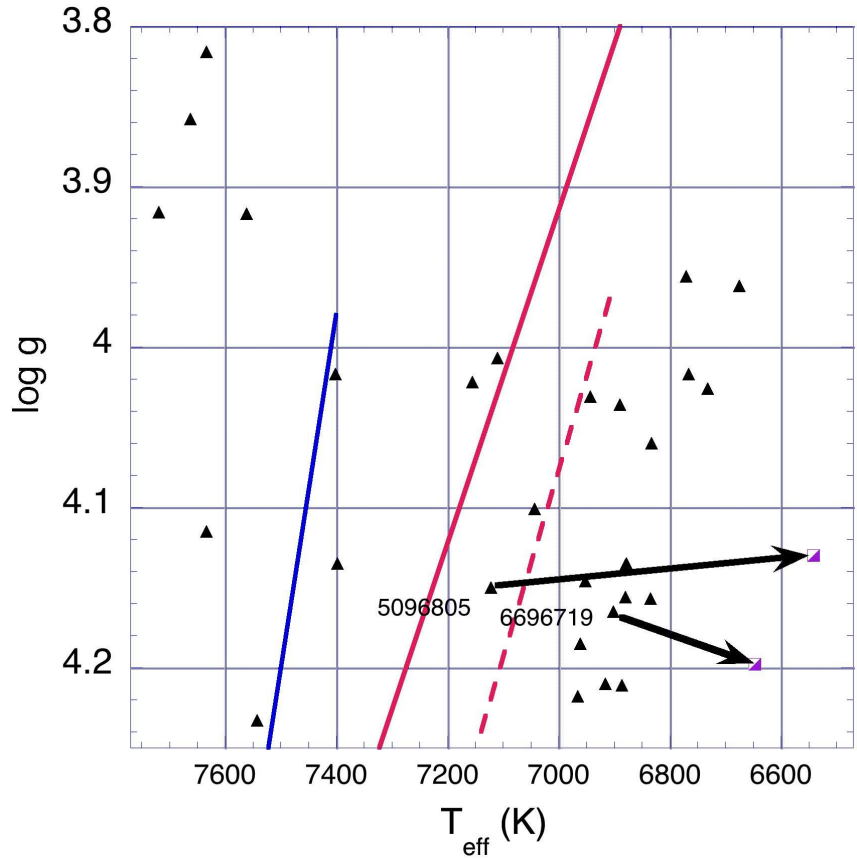}
\caption{Same as Fig. \ref{CSFig5a}, but without the +229 K offset applied to the KIC T$_{\rm eff}$, and with two stars labeled that have T$_{\rm eff}$ and log g determined by LAMOST spectra.  These stars would lie to the red of instability regions by several hundred K (purple arrows) if the LAMOST T$_{\rm eff}$ were used instead of the KIC T$_{\rm eff}$ value.}
\label{CSFig5b}
\end{figure*}

In Table \ref{table2} are highlighted the two `constant' stars near the red edge of the pulsation instability regions that also have LAMOST-derived T$_{\rm eff}$ and log g.  Figures \ref{CSFig4a} and \ref{CSFig4b} show the light curves and amplitude spectra of these stars.  The LAMOST T$_{\rm eff}$ is 582 K lower for KIC 5096805, and 395 K lower for KIC 6696719.  Figures \ref{CSFig5a} and \ref{CSFig5b} show a zoom-in on the pulsation instability regions with and without the +229 K T$_{\rm eff}$ offset.  Both of these stars would move to the red of the instability regions if the LAMOST-derived T$_{\rm eff}$ were to be adopted (see arrows in Fig. \ref{CSFig5b}).

\begin{figure*}%tb
\center
\includegraphics[width=1.5\columnwidth]{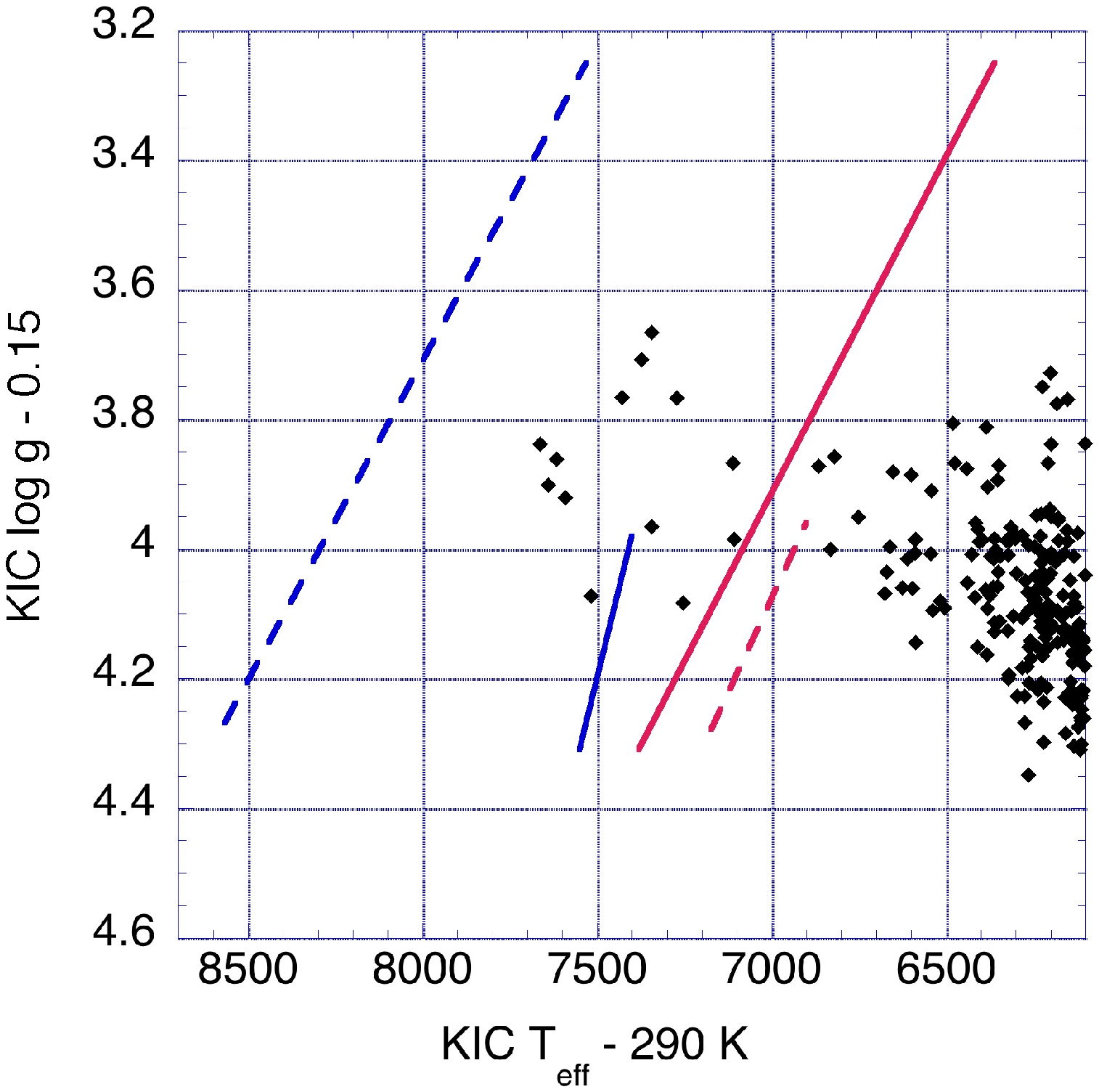}
\caption{Same as Fig. \ref{CSFig1b}, but with -290 K and -0.15 dex applied to KIC T$_{\rm eff}$ and log g.  In this plot, 15 stars lie within the pre-{\it Kepler} pulsation instability regions.}
\label{CSFig6a}
\end{figure*}

\begin{figure*}%tb
\center
\includegraphics[width=1.5\columnwidth]{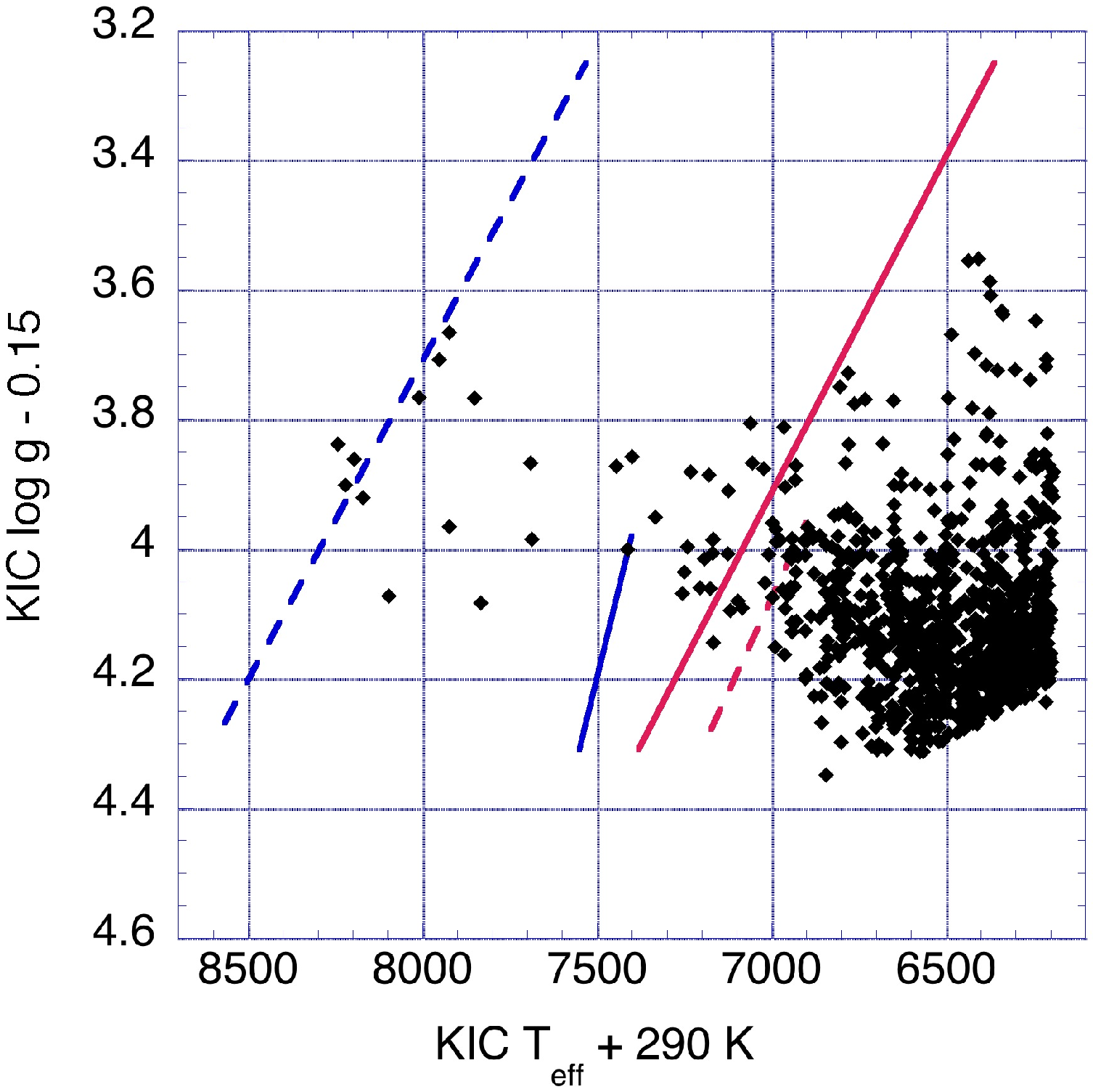}
\caption{Same as Fig. \ref{CSFig1b}, but with +290 K and -0.15 dex applied to KIC T$_{\rm eff}$ and log g.  In this plot, 52 stars lie within the pre-{\it Kepler} pulsation instability regions.}
\label{CSFig6b}
\end{figure*}

It is possible that the spectroscopically derived T$_{\rm eff}$ for other `constant' stars would also be lower, moving them out of the pulsation instability regions.  However, for some of the `constant' stars in the center or toward the blue of the $\delta$ Sct instability regions, T$_{\rm eff}$ changes to the KIC values greater than the average errors would be required to move them entirely out of the instability regions.  Figures \ref{CSFig6a} and \ref{CSFig6b} show the `constant' star sample of Fig. \ref{CSFig1b} (without the +229 K offset), and with the sample stars moved by an average (random) T$_{\rm eff}$ error of $\pm$ 290 K taken from \citet{uytterhoeven11}.  We also moved all of the stars in log g by -0.15 dex to roughly take into account the systematically higher average log g in the KIC catalog values vs. the LAMOST log g for log g $>$ 4.2, and the KIC overestimation bias found by \citet{verner11}.  In the first case (Fig. \ref{CSFig6a}), 15 sample stars remain within the ground-based pre-{\it Kepler} pulsation instability regions, and in the second case (Fig. \ref{CSFig6b}), 52 stars remain.  Therefore it is likely that, if the KIC T$_{\rm eff}$ errors are random, at least some `constant' stars that should show pulsations remain to be explained.

\clearpage
\onecolumn
\begin{deluxetable}{ccccccccc}
\tablecolumns{9} 
%\tablewidth{} 
\tablecaption{Spectral type, T$_{\rm eff}$ and log g determined from LAMOST spectra for 54 stars in common with our sample.}
\tablehead{\colhead{KIC}        &
           \colhead{K$_p$}                & 
           \colhead{LAMOST}      &
           \colhead{KIC}               &
           \colhead{LAMOST }     &    
           \colhead{LAMOST }     &                         
           \colhead{KIC }              &
           \colhead{LAMOST}      &        
           \colhead{LAMOST}    \\
           \colhead{Number}         &
           \colhead{Mag}               &               
           \colhead{Spectral Type}    &            
           \colhead{T$_{\rm eff}$}                    & 
           \colhead{T$_{\rm eff}$}                    & 
           \colhead{T$_{\rm eff}$ error}            & 
            \colhead{log g\tablenotemark{c}}                 & 
            \colhead{log g}                 & 
            \colhead{log g error}       }
\startdata
4374006  &  14.407  &  F5V  &  6071  &  5851  &  137  &  4.399  &  4.22  &  0.14  \\  
4561941  &  14.552  &  F2V  &  6348  &  6268  &  213  &  3.772  &  4.13  &  0.12  \\  
4568087  &  14.781  &  F0IV  &  7153  &  6898  &  121  &  4.127  &  4.1  &  0.11  \\  
4735997  &  14.716  &  F4V  &  6048  &  6085  &  211  &  4.135  &  4.19  &  0.11  \\  
4929320  &  14.496  &  G0V  &  5914  &  5856  &  121  &  4.318  &  4.14  &  0.21  \\  
4995588  &  14.961  &  A5IV  &  7401  &  7700  &  156  &  4.16  &  3.86  &  0.14  \\  
4999718  &  14.444  &  F5  &  6331  &  6202  &  98  &  4.405  &  4.19  &  0.12  \\  
5088361  &  15.109  &  F5  &  6299  &  6050  &  129  &  4.289  &  4.2  &  0.11  \\  
5096460  &  14.135  &  F5V  &  6826  &  6569  &  132  &  4.1  &  4.14  &  0.11  \\  
5096805\tablenotemark{b}   &  14.510  &  F4V  &  7122  &  6541  &  128  &  4.15  &  4.13  &  0.12  \\  
5168478  &  14.473  &  F3/F5V  &  6211  &  6397  &  81  &  4.174  &  4.17  &  0.1  \\  
5257286  &  14.326  &  F5V  &  6429  &  5855  &  84  &  4.483  &  4.17  &  0.11  \\  
5342935  &  14.033  &  F5V  &  6558  &  6595  &  128  &  4.36  &  4.14  &  0.11  \\  
5437797  &  14.370  &  F8V  &  5971  &  5738  &  71  &  4.26  &  4.2  &  0.15  \\  
5610296  &  14.846  &  F5  &  6348  &  6078  &  128  &  4.429  &  4.21  &  0.12  \\  
5629082  &  14.026  &  G0V  &  5967  &  5929  &  120  &  4.283  &  4.15  &  0.13  \\  
5629366\tablenotemark{a}  &  14.002  &  G0V  &  6012  &  5904  &  141  &  3.873  &  4.11  &  0.17  \\  
5629366  &  14.002  &  G0V  &   6012  &  5928  &  118  & 3.873  &  4.12  &  0.16  \\  
5685860  &  14.385  &  F5  &  6022  &  6120  &  131  &  4.271  &  4.21  &  0.12  \\  
5716288  &  14.301  &  F6IV  &  6398  &  6367  &  142  &  4.453  &  4.09  &  0.15  \\  
5783368  &  14.806  &  A6V  &  7910  &  7958  &  107  &  3.835  &  3.96  &  0.11  \\  
5809732  &  14.098  &  F1V  &  6613  &  6891  &  156  &  4.122  &  4.09  &  0.11  \\  
5858178  &  14.416  &  F5  &  6088  &  5971  &  136  &  4.325  &  4.19  &  0.12  \\  
6113533  &  14.409  &  F8V  &  6081  &  6019  &  106  &  4.355  &  4.13  &  0.13  \\  
6132246  &  14.462  &  F4V  &  6539  &  6504  &  89  &  4.227  &  4.12  &  0.11  \\  
6207546  &  14.321  &  F5V  &  7134  &  6569  &  142  &  4.305  &  4.14  &  0.11  \\  
6292725  &  14.168  &  F7III  &  5905  &  5776  &  184  &  4.141  &  4.14  &  0.26  \\  
6446706  &  14.826  &  F6V  &  6217  &  6393  &  167  &  4.329  &  4.12  &  0.13  \\  
6446775  &  14.348  &  G0V  &  6133  &  6004  &  110  &  4.383  &  4.19  &  0.13  \\  
6452044  &  14.218  &  F2V  &  6510  &  6386  &  82  &  4.23  &  4.13  &  0.12  \\  
6526666  &  14.176  &  F0V  &  6832  &  6504  &  99  &  4.219  &  4.12  &  0.11  \\  
6696719\tablenotemark{b}  &  14.274  &  F3V  &  6902  &  6507  &  84  &  4.165  &  4.13  &  0.11  \\  
7038888  &  14.316  &  F0V  &  6812  &  6647  &  106  &  4.198  &  4.14  &  0.11  \\  
7191541  &  14.303  &  F7IV  &  6633  &  6317  &  98  &  4.25  &  4.1  &  0.13  \\  
7205852  &  14.393  &  F5  &  6253  &  6250  &  105  &  4.352  &  4.13  &  0.12  \\  
7439291  &  14.329  &  F5  &  6380  &  6204  &  90  &  4.212  &  4.15  &  0.12  \\  
7448387  &  14.716  &  G5V  &  6315  &  5817  &  110  &  4.199  &  4.11  &  0.18  \\  
7729149  &  14.024  &  F5V  &  6442  &  6394  &  74  &  4.273  &  4.11  &  0.11  \\  
7746728  &  14.446  &  F8V  &  6081  &  6077  &  115  &  4.379  &  4.08  &  0.13  \\  
7748881  &  14.143  &  F6V  &  6367  &  6242  &  131  &  4.262  &  4.19  &  0.11  \\  
8082314  &  14.299  &  G5  &  5917  &  5819  &  95  &  4.206  &  4.24  &  0.14  \\  
8158284  &  14.415  &  G5  &  5995  &  5858  &  89  &  4.36  &  4.19  &  0.14  \\  
8316032  &  14.511  &  F3V  &  6250  &  6467  &  108  &  3.979  &  4.14  &  0.11  \\  
8327183  &  14.080  &  F7IV  &  6021  &  6333  &  109  &  4.198  &  4.13  &  0.12  \\  
8361689  &  14.647  &  F8V  &  6201  &  6215  &  75  &  4.42  &  4.13  &  0.13  \\  
8414160  &  14.193  &  G0V  &  6491  &  6221  &  95  &  4.228  &  4.14  &  0.12  \\  
8415383  &  14.010  &  F1V  &  7165  &  6957  &  137  &  4.06  &  4.07  &  0.12  \\  
8626053  &  14.384  &  F4V  &  6217  &  6024  &  145  &  4.306  &  4.12  &  0.17  \\  
8767298  &  14.558  &  F5V  &  6442  &  6500  &  221  &  4.475  &  4.09  &  0.13  \\ 
8773083  &  14.000  &  F6V  &  5931  &  6216  &  91  &  4.105  &  4.11  &  0.12  \\  
9238849  &  14.296  &  F5V  &  6056  &  6513  &  100  &  4.122  &  4.13  &  0.11  \\  
9479588\tablenotemark{a}  &  14.723  &  F7IV  &  6141  &  6295  &  106  &  4.304  &  4.13  &  0.11  \\  
9479588  &  14.723  &  F5  &   6141   &  6331  &  103  &  4.304   &  4.14  &  0.12  \\  
9724292  &  14.397  &  F0III  &  6807  &  7215  &  157  &  4.07  &  3.8  &  0.14  \\  
9905069  &  14.582  &  F6IV  &  6498  &  6123  &  109  &  4.202  &  4.09  &  0.14  \\  
9905074\tablenotemark{a}  &  14.570  &  F5V  &  6424  &  6315  &  122  &  4.454  &  4.15  &  0.13  \\  
9905074  &  14.570  &  F5  &  6424  &  6331  &  103  &   4.454   &  4.14  &  0.12 \\
\enddata
\tablenotetext{a}{Three stars have two LAMOST spectra taken months apart that give different T$_{\rm eff}$/log g interpretation.}
\tablenotetext{b}{Two stars are among those on our list of `constant' stars within the pulsation instability region seen in Figs. \ref{CSFig1a} and \ref{CSFig1b}. The LAMOST-determined T$_{\rm eff}$ are significantly lower for these two stars and would place them to the red of the instability regions (Fig. \ref{CSFig5b}).}
\tablenotetext{c}{The KIC gives log g to three digits after the decimal, and so these digits were retained even though the log g value is not likely to be this accurate.}
\label{table2}
\end{deluxetable}
\twocolumn

\section{Super-Nyquist frequencies}

Among the list of explanations for the `constant' stars given in Paper 1 is the possibility that the `constant' stars are pulsating at frequencies higher than the 24.4 c/d Nyquist frequency for long-cadence {\it Kepler} data.  However, \citet{murphy13} show that super-Nyquist frequencies do appear in Fourier transforms of the long-cadence data at a lower frequency reflected below the Nyquist frequency.  If the data set is long enough (of order 1 year), the Nyquist reflections may even be distinguishable from real frequencies via their side-lobes, and therefore the true super-Nyquist frequency may be determined.

\begin{figure*}%tb
\center
\includegraphics[width=1.5\columnwidth]{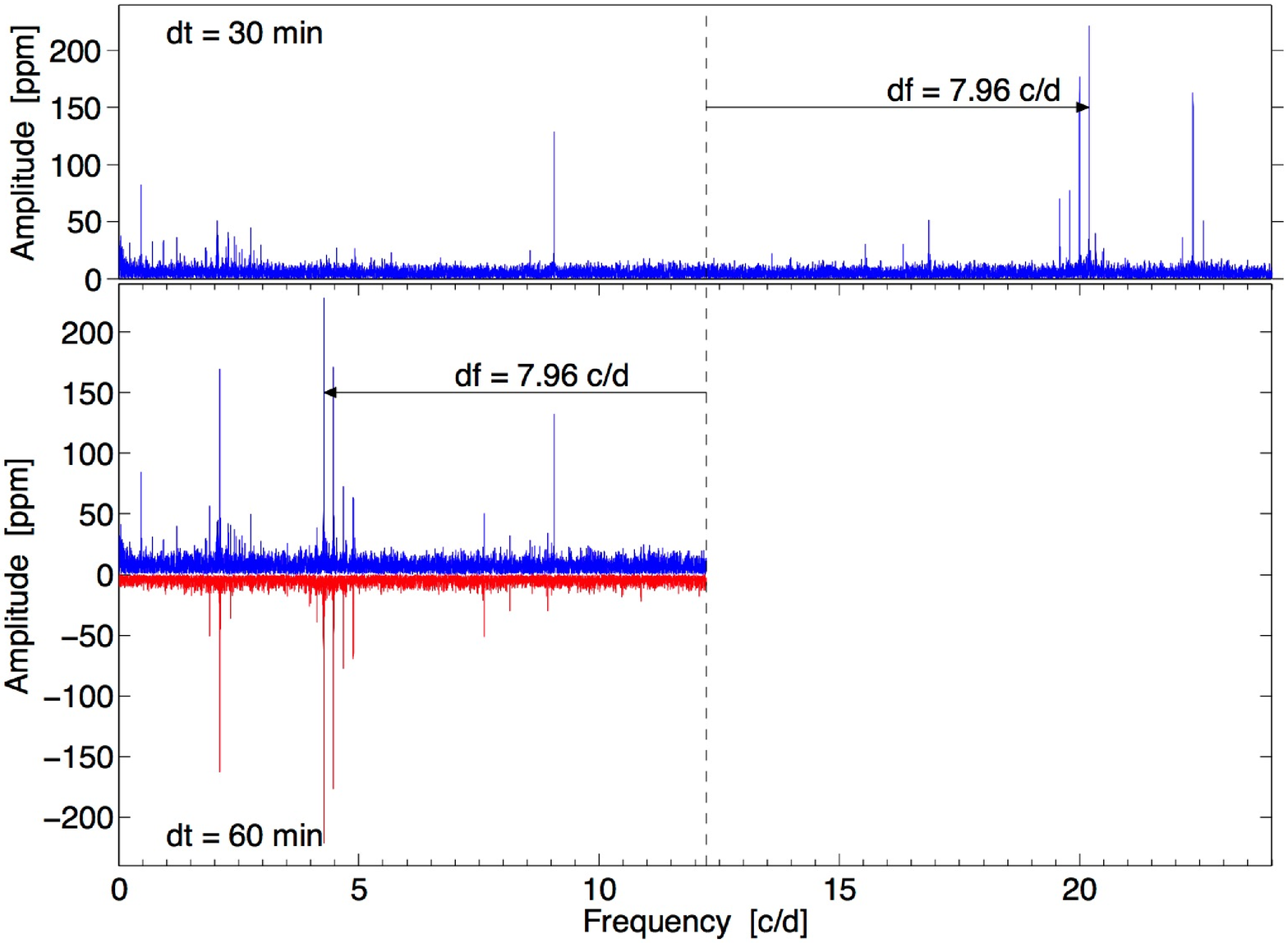}
\caption{Top panel:  Amplitude spectrum from Fourier transform of light curve for {\it Kepler} B star taken in long cadence, with Nyquist frequency 24.4 c/d. Bottom panel:  Amplitude spectrum from Fourier transform of same light curve sub-sampled at 1-hour cadence, with Nyquist frequency 12.2 c/d.  The super-Nyquist frequencies seen on the top panel are ÔmirroredÕ at low frequency. The amplitude spectrum above 12.2 c/d from the top panel is inverted and reflected (red) to compare with the spectrum in the sub-sampled light curve.}
\label{CSFig7}
\end{figure*}

To illustrate this effect, we show in Fig.~\ref{CSFig7} the amplitude spectrum for a {\it Kepler} B star observed in long cadence ($\sim$30 min integration per data point, top panel), with Nyquist frequency 24.4 c/d.  Some high-frequency peaks are evident at about 20 c/d.   The bottom panel shows the amplitude spectrum obtained using the light curve sampled every $\sim$1 hour, so the Nyquist frequency is 12.2 c/d (vertical dashed line).  The frequencies above the 24.4 c/d Nyquist frequency from the first panel can be seen ÔmirroredÕ below the 12.2 c/d Nyquist frequency.  The arrow highlights the highest amplitude peak at 7.96 c/d above the Nyquist frequency that shows up at 7.96 c/d below the Nyquist frequency for the sub-sampled set. The red curve shows the amplitude spectrum above 12.2 c/d from the top panel inverted and reflected to compare with the spectrum in the sub-sampled light curve.

We conclude that it is unlikely that any of our `constant' stars have pulsation frequencies above the Nyquist frequency 24.4 c/d at amplitudes typical of $\gamma$ Dor or $\delta$ Sct pulsations ($>$100 ppm), as their reflections should appear at lower frequencies in the Fourier transform.  If this were the case, a time series of at least a year of long-cadence data would be needed to distinguish a Nyquist reflection frequency from a true frequency, as the barycentric time corrections to the {\it Kepler} data then begin to create detectable sidelobe peaks around the super-Nyquist frequency peaks \citep{murphy2013}.

\section{Remaining explanations for `constant' stars}

While we have discussed two of the five explanations suggested for the `constant' stars in Paper 1, we have not addressed three others.  First, the stars may be pulsating in higher spherical harmonic degree modes (e.g., degree $>$ 0-3) that are not easily detectable photometrically. Such high-degree modes may be detectable spectroscopically, but high resolution and time series spectra for such faint stars require continuous observation for weeks with 10-meter class telescopes, and is impossible to organize.  It is interesting that for a few brighter $\gamma$ Dor stars where multisite spectroscopic data has been collected and analyzed, frequencies first detected photometrically are also seen spectroscopically via their line-profile variations \citep[e.g., HD 12901,][]{brunsden12}, while for other $\gamma$ Dor stars, there seems to be almost no relationship between frequencies detected spectroscopically, and those found photometrically \citep[e.g., HD 49434,][]{brunsden14a, brunsden14b}.

Second, as can be seen from some of the example constant-star power spectra in Figs. \ref{CSFig2a} through \ref{CSFig2f}, there may be pulsation modes with amplitudes at or below the noise level of this data; the modes could possibly be detected with a longer time series or by reducing the noise levels, possibly making use of the {\it Kepler} pixel data.  Third, a physical mechanism may be operating, e.g., diffusive helium or element settling, that inhibits pulsations for some stars.

\citet{murphy14} identified about 50 bright (V magnitude $\le$ 10), non chemically peculiar stars in or near the $\delta$ Sct instability region observed by {\it Kepler} that showed no variability at $\delta$ Sct p-mode frequencies of 5 to 50 c/d with amplitude higher than 50 $\mu$mag.  They used high-resolution spectroscopic observations to derive a more accurate T$_{\rm eff}$ and log g, as well as $v$~sin~$i$ and metallicity.  They find that many of their non-pulsating stars could plausibly lie outside the $\delta$ Sct instability region due to remaining uncertainties in T$_{\rm eff}$ or if the star was actually a binary and misinterpreted as a single star.  However, they find one non-pulsating chemically normal star in the center of the $\delta$ Sct pulsation region that remains to be explained.

\section{Distribution of `non-constant' stars}

\begin{figure*}%tb
\center
\includegraphics[width=1.5\columnwidth]{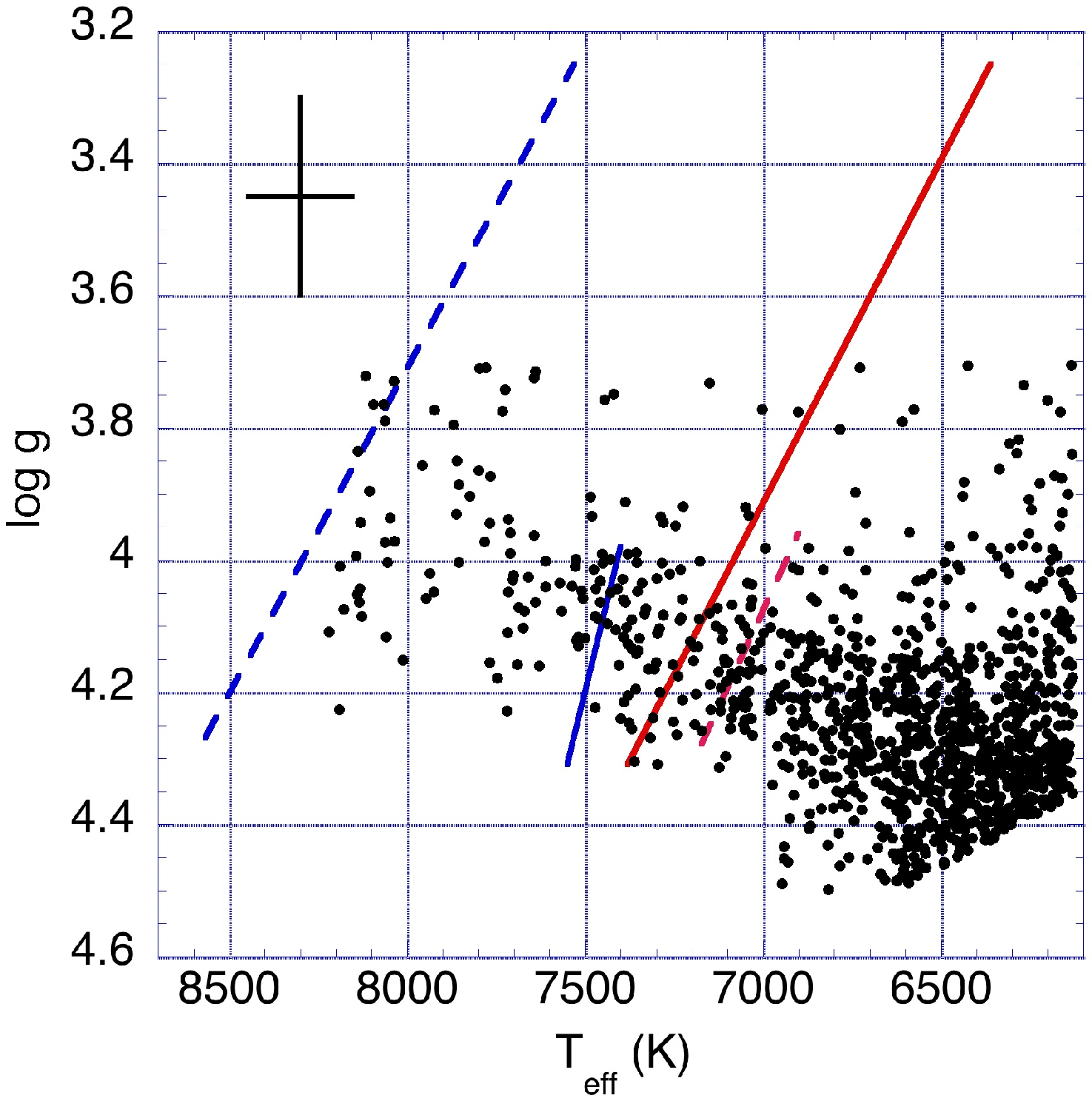}
\caption{Location of 1147 sample stars that are not `constant' (see definition in text) in the log g--T$_{\rm eff}$ diagram, along with the $\delta$ Sct (dashed lines) and $\gamma$ Dor (solid lines) instability strip boundaries established from pre-{\it Kepler} ground-based observations \citep{rodriguezbreger01, handlershobbrook02}.  The T$_{\rm eff}$ of the stars has been shifted by +229 K to account for the systematic offset in temperatures of the {\it Kepler} Input Catalog and SDSS photometry for this temperature range as determined by \citet{pinsonneault12}.  The black cross shows an error bar on log g (0.3 dex) and T$_{\rm eff}$ (290 K) determined by comparisons of KIC values and values derived from ground-based spectroscopy for brighter {\it Kepler} targets \citep{uytterhoeven11}. Most of these stars have light curves consistent with stellar activity (spots) or binarity.}
\label{CSFig8a}
\end{figure*}

\begin{figure*}%tb
\center
\includegraphics[width=1.5\columnwidth]{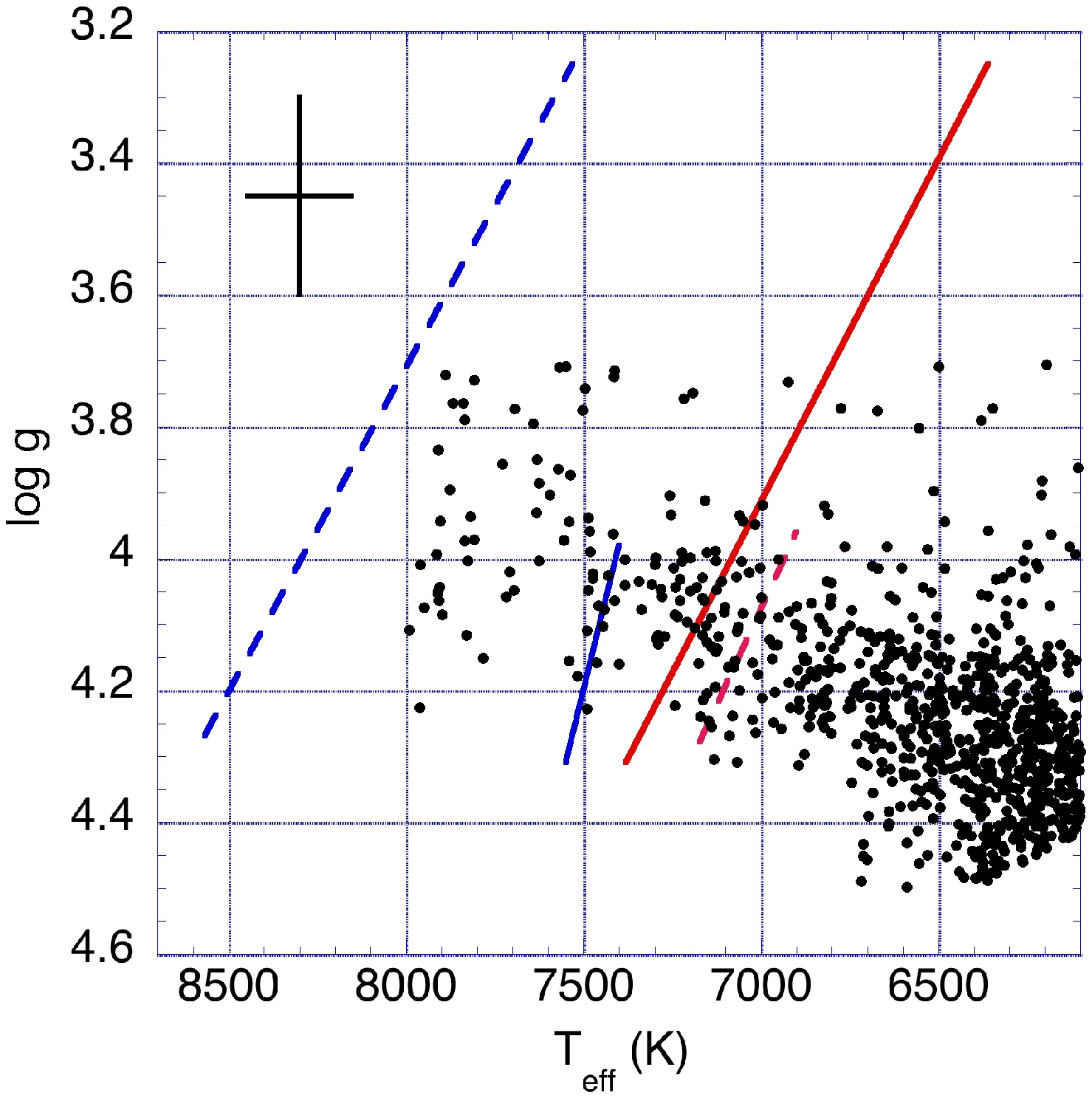}
\caption{Same as Fig. \ref{CSFig8a} without the +229 K offset applied to the KIC T$_{\rm eff}$.}
\label{CSFig8b}
\end{figure*}

Figures \ref{CSFig8a} and \ref{CSFig8b} show how the remaining 1147 Ônon-constantÕ stars are distributed in relation to the instability regions established from ground-based observations, with and without the +229 K offset applied to the KIC T$_{\rm eff}$.  \citet{bradley14, bradley15} discuss and categorize the Ônon-constantÕ stars in the combined set of stars observed in Q6-13 and Q14-17.  The light curves of many of these stars show irregular low-frequency variations consistent with rotating spots or stellar activity; many are eclipsing binaries.  \citet{balona11, balona13, balona14}, \citet{balonaguzik11} and \citet{balonadziembowski11} discuss the light-curve properties of many of the A-F stars observed by {\it Kepler}.  As discussed by \citet{bradley14, bradley15} and as also found by \citet{uytterhoeven11} and \citet{grigahcene10}, $\delta$ Sct and $\gamma$ Dor candidates lie outside the pulsation instability regions established by ground-based observations.  Additional pulsation driving mechanisms may be needed to explain the observed frequencies.   A new pulsation mechanism involving fluctuations in turbulent pressure in the hydrogen-ionization region, for example, has been proposed to explain pulsations at $\delta$ Sct-like frequencies by \citet{antoci14}.

\section{Conclusions and future work}

In Paper I discussing a sample of 633 A-F main-sequence stars observed by {\it Kepler} during Quarters 6-13, 359 stars, or roughly 60\%, were found to be `constant', defined as showing no frequencies above 20 ppm in their light curves between 0.2 and 24.4 c/d.  In that sample, we found very few (only six) photometrically non-varying stars within the $\gamma$ Dor and $\delta$ Sct pulsation instability regions established by pre-{\it Kepler} ground-based observations.   Here, in our sample of 2137 stars observed by {\it Kepler} in Quarters 14-17, 990 stars, or 46\% were found to be `constant', with 34 of these stars lying within the pulsation instability regions.  The percentage of `constant' stars is lower because the targets were selected by cross-correlating with the list of stars that showed variability in Q0 full-frame images.  

We compare the KIC T$_{\rm eff}$ with the T$_{\rm eff}$ derived for 54 stars in common with our GO targets with T$_{\rm eff}$ derived from LAMOST spectroscopy.  While there is significant scatter, on average the LAMOST derived T$_{\rm eff}$ is the same or slightly smaller than the KIC T$_{\rm eff}$; therefore, it is questionable whether we should have applied a +229 K increase to the KIC T$_{\rm eff}$ for stars of this spectral type, as suggested by the analysis of \citet{pinsonneault12} using Sloan multicolor photometry.  Without the offset, only 17 stars lie within the instability region.  For two `constant' stars with LAMOST observations, adopting the LAMOST-derived T$_{\rm eff}$ moves them to the red of the instability regions.  A likely explanation for many of the `constant' stars within the instability region is therefore that their KIC T$_{\rm eff}$ and/or log g is not correct.  However, for some of the hottest constant stars, the T$_{\rm eff}$ adjustment from KIC would be quite large; in addition, T$_{\rm eff}$ or log g adjustments may also move some `constant' stars that are outside of the instability regions into the instability regions, so there may be many `constant' stars that should be pulsating that remain to be explained. 

It is also unlikely that `constant' stars actually have undetected frequencies above the long-cadence Nyquist frequency.  As \citet{murphy13} show, and as we also illustrate, such frequencies would be mirrored below the Nyquist frequency and be visible unless the amplitudes are low compared to the noise level.

For future work, we hope to make use of LAMOST data to improve T$_{\rm eff}$, log g, and metallicity determinations for these faint stars, and model some of the `constant' stars that may still lie within the instability regions.  We also could use the {\it Kepler} pixel data or refined light curve analyses to reduce noise levels to establish the significance of modes  with very low amplitudes.  We would like to use stellar modeling to determine how much diffusive settling is possible and necessary to eliminate pulsations in $\gamma$ Dor or $\delta$ Sct stars.  We also hope to carry out model calculations to explore alternate pulsation driving mechanisms that may help explain the {\it Kepler} stars that show pulsations at unexpected frequencies or lie outside of predicted instability boundaries.  

\section*{Acknowledgments}
The authors are grateful for data and funding through the NASA {\it Kepler} Guest Observer Program Cycles 1-4.  KU acknowledges financial support by the Spanish National Plan of R\&D for 2010, project AYA2010-17803. This work has benefited from funding by the Project FP7-PEOPLE-IRSES:ASK no 269194. J.G. thanks S. Murphy and H. Shibahashi for useful discussions.


\begin{thebibliography}{}

\bibitem[Antoci et al.(2014)]{antoci14}Antoci, V., Cunha, M., Houdek, G., Kjeldsen, H., Trampedach, R., Handler, G., L\"{u}ftinger, T., Arentoft, T., \& Murphy, S., ``The role of turbulent pressure as a coherent pulsational driving mechanism: The case of the $\delta$ Scuti star HD 187547,'' \apj, 796, 118 (2014).
\bibitem[Balona et al.(1994)]{balona94}Balona, L.A., Krisciunas, K., \& Cousins, A.W.J., ``$\gamma$ Doradus:  Evidence for a new class of pulsating star,''\mnras,  270, 905 (1994).
\bibitem[Balona \& Dziembowski(1999)]{balonadziembowski99}Balona, L.A. \& Dziembowski, W.A., ``Excitation and visibility of high-degree modes in stars,'' \mnras,  309, 221 (1999).
\bibitem[Balona \& Dziembowski(2011)]{balonadziembowski11}Balona, L.A. \& Dziembowski, W.  ``{\it Kepler} Observations of $\delta$ Scuti Stars,'' \mnras, 417, 591 (2011).
\bibitem[Balona(2011)]{balona11}Balona, L.A., ``Rotational light variations in {\it Kepler} observations of A-type stars,'' \mnras, 415, 1691 (2011).
\bibitem[Balona et al.(2011)]{balonaguzik11}Balona, L.A., Guzik, J.A., Uytterhoeven, K., Smith, J.C., Tenenbaum, P., \& Twicken, J.D., ``The {\it Kepler} view of $\gamma$ Doradus stars,'' \mnras, 415, 3531 (2011).
\bibitem[Balona(2013)]{balona13}Balona, L.A., ``Activity in A-type stars,'' \mnras, 431, 2240 (2013).
\bibitem[Balona(2014)]{balona14}Balona, L.A., ``Low frequencies in {\it Kepler} $\delta$ Scuti stars,''  \mnras, 437, 1476 (2014).
\bibitem[Borucki et al.(2010)]{borucki10}Borucki, W.J. et al., ``{\it Kepler} planet-detection mission: Introduction and first results,'' Science, 327, 977 (2010).
\bibitem[Bradley et al.(2014)]{bradley14}Bradley, P.A., Guzik, J.A., Miles, L.F., Jackiewicz, J., Uytterhoeven, K., \& Kinemuchi, K., ``Analysis of $\gamma$ Doradus and $\delta$ Scuti Stars Observed by {\it Kepler},'' Proc. IAU Symposium 301:  Precision Asteroseismology, eds. J.A.~Guzik, W. Chaplin, G. Handler, \& A. Pigulski, Cambridge U. Press, pp. 387-388 (2014).
\bibitem[Bradley et al.(2015)]{bradley15}Bradley, P.A., Guzik, J.A., Miles, L.F., Jackiewicz, J., Uytterhoeven, K., \& Kinemuchi, K., ``Results of a search for $\gamma$ Doradus and $\delta$ Scuti stars with the {\it Kepler} Spacecraft,'' \aj, 149, 68 (2015).
\bibitem[Brunsden et al.(2012)]{brunsden12}Brunsden, E., Pollard, K. R., Cottrell, P. L., Wright, D. J., \& De Cat, P., ``Spectroscopic pulsational frequency identification and mode determination of $\gamma$ Doradus star HD 12901,'' \mnras, 427, 2512 (2012).
\bibitem[Brunsden et al.(2014a)]{brunsden14a}Brunsden, E., Pollard, K. R., Cottrell, P. L., Wright, D. J., De Cat, P., \& Kilmartin, P. M., ``Spectroscopy of $\gamma$ Dor Stars,'' Precision Asteroseismology, IAU Symposium 301, Wroclaw, Poland, August 2013, eds. Guzik, J.A., Chaplin, W.J., Handler, G., \& Pigulski, A., p. 113 (2014a).
\bibitem[Brunsden et al.(2014b)]{brunsden14b}Brunsden, E., Pollard, K. R., Cottrell, P. L., Uytterhoeven, K., Wright, D. J., \& De Cat, P., ``The classification of frequencies in the $\gamma$ Doradus / $\delta$ Scuti hybrid star HD 49434,''  \mnras, accepted Dec. 2014, 2014arXiv1412.2828B (2014b).
\bibitem[Chevalier et al.(1971))]{chevalier71}Chevalier, C., ``Short-period variables. VIII. Evolution and pulsation of $\delta$ Scuti stars'', \aap, 14, 24 (1971).
\bibitem[Dupret et al.(2004)]{dupret04}Dupret, M.A., Grigahc{\'e}ne, A., Garrido, R., Gabriel, M., \& Scuflaire, R., ``Theoretical instability strips for $\delta$ Scuti and $\gamma$ Doradus stars,'' \aap,  414, L17 (2004).
\bibitem[Dupret et al.(2005)]{dupret05}Dupret, M.A., et al.,``Convection-pulsation coupling. II. Excitation and stabilization mechanisms in $\delta$ Sct and $\gamma$ Dor stars,'' \aap, 435, 927 (2005).
\bibitem[Gaulme \& Guzik(2014)]{gaulme14}Gaulme, P. \& Guzik, J.A.,  ``Searching for pulsations in {\it Kepler} eclipsing binary stars,'' Proc. IAU Symposium 301:  Precision Asteroseismology, eds. J.A. Guzik, W. Chaplin, G. Handler, \& A. Pigulski, Cambridge U. Press, pp. 413-414 (2014).
\bibitem[Gilliland et al.(2010)]{gilliland10}Gilliland, R. et al., ``{\it Kepler} asteroseismology program:  Introduction and first results,'' \pasp, 122, 131 (2010). 
\bibitem[Grigahc{\'e}ne et al.(2006)]{grigahcene06}Grigahc{\'e}ne, A., Dupret., M.-A., Garrido, R., and Gabriel, M., ``Influence of overshooting and metallicity on the $\delta$ Scuti and $\gamma$ Doradus instability strips,'' Mem. Sa.A.It., 7, 129 (2006).
\bibitem[Grigahc{\'e}ne et al.(2010)]{grigahcene10}Grigahc{\'e}ne, A., et al., ``Hybrid $\gamma$ Doradus--$\delta$ Scuti pulsators:  New Insights into the physics of the sscillations from {\it Kepler} observations,'' \apj~(Letters), 713L, 192 (2010).
\bibitem[Guzik et al.(2000)]{guzik00}Guzik, J.A., Kaye, A.B, Bradley, P.A., Cox, A.N., \& Neuforge, C., ``Driving the gravity-mode pulsations in $\gamma$ Doradus variables,'' \apj~(Letters), 542, L57 (2000).
\bibitem[Guzik et al.(2013)]{guzik13}Guzik, J.A., Bradley, P.A., Jackiewicz, J., Uytterhoeven, K., \& Kinemuchi, K., ``The Occurrence of non-pulsating stars in the $\gamma$ Doradus/$\delta$ Scuti pulsation instability region,'' Astronomical Review, Vol. 8.4, pp. 4-30; arXiv:1403.8013 (Paper 1, 2013).
\bibitem[Guzik et al.(2014)]{guzik14}Guzik, J.A., Bradley, P.A., Jackiewicz, J., Uytterhoeven, K., \& Kinemuchi, K., ``The occurrence of non-pulsating stars in the $\gamma$ Doradus and $\delta$ Scuti pulsation instability regions,'' Proc. IAU Symposium 301:  Precision Asteroseismology, eds. J.A. Guzik, W. Chaplin, G. Handler, \& A. Pigulski, Cambridge U. Press, pp. 63-66 (2014).
\bibitem[Handler \& Shobbrook(2002)]{handlershobbrook02}Handler, G. \& Shobbrook, R.R.,``On the relationship between the $\delta$ Scuti and $\gamma$ Doradus pulsators,'' \mnras, 333, 251 (2002).
\bibitem[Henry et al.(2007)]{henry07}Henry, G., et al. ``Photometry and spectroscopy of 11 $\gamma$ Doradus stars,'' \aj, 133, 1421 (2007).
\bibitem[Houdek et al.(1999)]{houdek99}Houdek, G., Balmforth, N.J., Christensen-Dalsgaard, J., \& Gough, D.O., ``Amplitudes of stochastically excited oscillations in main sequence stars,'' \aap, 351, 582 (1999).
\bibitem[Jenkins et al.(2010)]{jenkins10}Jenkins, J.M., et al., ``Overview of the {\it Kepler} science processing pipeline,'' \apj~(Letters), 713L, 87 (2010).
\bibitem[Kaye et al.(1999)]{kaye99}Kaye, A.B., Handler, G., Krisciunas, K., Poretti, E. \& Zerbi, F.M., ``$\gamma$ Doradus stars:  Defining a new class of pulsating variables,'' \pasp, 111, 840 (1999).
\bibitem[Kinemuchi et al.(2011)]{kinemuchi11}Kinemuchi, K., et al., ``{\it Kepler} Full-Frame Image Variable Star Catalog,''  \baas, Vol. 43, 21720106 (2011).
\bibitem[Latham et al.(2005)]{latham05}Latham, D.W., Brown, T.M., Monet, D.G., Everett, M., Esquerdo, G. A., \& Hergenrother, C. W.  ``The {\it Kepler} Input Catalog,'' Bulletin of the American Astronomical Society Meeting 207, 37, 1340 (2005).
\bibitem[McNamara et al.(2012)]{mcnamara12}McNamara, B., Jackiewicz, J., \& McKeever, J., ``The Classification of {\it Kepler} B-Star Variables,'' \aj,  143, 101 (2012).
\bibitem[Molenda-Zakowicz et al.(2013)]{zakowicz13} Molenda-Zakowicz, J., et al., ``Atmospheric parameters of 169 F-, G-, K- and M-type stars in the {\it Kepler} field,'' \mnras, 434, 1422 (2013).
\bibitem[Molenda-Zakowicz et al.(2014)]{zakowicz14} Molenda-Zakowicz, J., De Cat, P., Fu, Jian-Ning, Yang, Xiao-Hu, and the LAMOST-{\it Kepler} collaboration, ``LAMOST observations in the {\it Kepler} field,'' Precision Asteroseismology, IAU Symposium 301, Wroclaw, Poland, August 2013, eds. Guzik, J.A., Chaplin, W.J., Handler, G., \& Pigulski, A., p. 457 (2014).
\bibitem[Murphy, Shibahashi, \& Kurtz(2013)]{murphy13}Murphy, S., J., Shibahashi, H., \& Kurtz, D.W., ``SuperNyquist asteroseismology with the {\it Kepler} Space Telescope,'' \mnras, 430, 2896 (2013).
\bibitem[Murphy et al.(2014)]{murphy14}Murphy, S., Bedding, T.J., Niemszura, E., Kurtz, D.W., \& Smalley, B., ``A search for non-pulsating, chemically normal stars in the $\delta$ Scuti instability strip using {\it Kepler} data,'' \mnras, accepted Dec. 2014, 2014arXiv1412.7543M (2014).
\bibitem[Pinsonneault et al.(2012)]{pinsonneault12}Pinsonneault, M.H., An, D., Molenda-Zakowicz, J., Chaplin, W.J., Metcalfe, T.S., \& Bruntt, H.,  ``A revised effective temperature scale for the {\it Kepler} Input Catalog,'' \apjs, 199, 30 (2012).
\bibitem[Pinsonneault et al.(2013)]{pinsonneault13}Pinsonneault, M.H., An, D., Molenda-Zakowicz, J., Chaplin, W.J., Metcalfe, T.S., \& Bruntt, H.,  ``Erratum: A revised effective temperature scale for the {\it Kepler} Input Catalog,''\apjs, 208, 12 (2013).
\bibitem[Rodriguez et al.(2000)]{rodriguez00}Rodriguez, E., et al., ``A revised catalog of $\delta$ Scuti stars,''  \aap~Supplement Series, 144, 3 (2000).
\bibitem[Rodriguez \& Breger(2001)]{rodriguezbreger01}Rodriguez, E. \& Breger, M., ``$\delta$ Scuti and related stars: Analysis of the R00 Catalogue,'' \aap,  366, 178 (2001).
\bibitem[Turcotte et al.(1998)]{turcotte98}Turcotte, S., Richer, J., \& Michaud, G., ``Consistent evolution of F stars:  Diffusion, radiative accelerations, and abundance anomalies,'' \apj, 504, 559 (1998).
\bibitem[Uytterhoeven et al.(2011)]{uytterhoeven11}Uytterhoeven, K., et al., ``The {\it Kepler} characterization of the variability among A- and F-type stars,'' \aap, 534, A125 (2011).
\bibitem[Verner et al.(2011)]{verner11}Verner, G.A., et al., ``Verification of the {\it Kepler} Input Catalog from Asteroseismology of Solar-Type stars,''  \apj~(Letters), 738, L28 (2011).


\end{thebibliography}
\end{document}